\documentclass[12pt]{article}
\usepackage{amsmath}
\usepackage{amssymb}
\usepackage{amsthm}
\usepackage{psfrag}
\usepackage{graphicx}
\usepackage{hyperref}
\usepackage[usenames]{color}
\DeclareSymbolFont{matha}{OML}{txmi}{m}{it}
\DeclareMathSymbol{\varv}{\mathord}{matha}{118}

\newcommand\bw{\begin{widetext}}
\newcommand\ew{\end{widetext}}
\newcommand{\be}{\begin{equation}}
\newcommand{\ee}{\end{equation}}
\newcommand{\beqa}{\begin{eqnarray}}
\newcommand{\eeqa}{\end{eqnarray}}

\newcommand{\pd}{\partial}
\newcommand\m{\mu}

\newcommand\n{\nu}

\newcommand\s{\sigma}
\renewcommand\a{\alpha}
\renewcommand\b{\beta}
\renewcommand\l{\lambda}

\newcommand{\Gc}{G_{\text{cos}}}
\newcommand{\nn}{\nonumber}

\def\d{\partial}
\newcommand{\bseq}{\begin{subequations}}
\newcommand{\eseq}{\end{subequations}}

\renewcommand{\ln}{\mathop{\rm ln}\nolimits}

\begin{document}

\title{\vspace{-2cm} 
{\normalsize
\flushright CERN-PH-TH-2014-193 \\}
\vspace{0.6cm} 
\sc Phenomenology of
 theories of gravity without Lorentz invariance: \\
 the preferred frame case\footnote{Invited contribution to the special issue \emph{``Modified Gravity and Effects of Lorentz Violation"}
 to appear in  IJMPD.}}

\author{Diego Blas$^\flat$\footnote{diego.blas@cern.ch},  Eugene Lim$^\sharp$\footnote{eugene.a.lim@gmail.com}\\[2mm]
{\normalsize\it $^\flat$ CERN, Theory Division, 1211 Geneva, Switzerland.}\\[1mm]
{\normalsize\it $^\sharp$ Theoretical Particle Physics and Cosmology Group,}\\[-2mm]
 {\normalsize\it Physics Department,
Kings College London, Strand, London WC2R 2LS, United Kingdom }}
\maketitle

\begin{abstract}
Theories of gravitation without Lorentz invariance are  candidates of low-energy
descriptions of quantum gravity. In this review we will describe the phenomenological consequences
of the candidates associated to the existence of a preferred time direction. 
\end{abstract}

\newpage
\tableofcontents

\section{Introduction}\label{sec:introd}

Lorentz invariance (LI) is one of the cornerstones of our current understanding of gravitation. The privileged role of LI  was first established for electromagnetism, and later extended to all the physical phenomena. For gravitation, LI was essential for Einstein to find alternatives to Newton's theory, which eventually lead to general relativity (GR). The success of GR to describe all observed gravitational phenomena, together with its intrinsic mathematical elegance consolidated it as \emph{the theory} of gravitation. This was also interpreted  as a further proof of the 
fundamental importance of LI. 

However, this triumph is thwarted by the consequences of using GR to describe situations at very short distances. In fact, the quantum aspects of GR at energies beyond the Planck mass $M_P\approx 10^{19} \, \mathrm{GeV}$ and its generic predictions
of classical singularities present a challenge that calls for a more compete theory\footnote{The necessity of a small cosmological constant
and the small tensions of the standard cosmological model  $\Lambda$CDM with observations are sometimes used as extra motivation to modify GR.}.  It is not our purpose to review the different proposals explored so far. We will just notice that those preserving LI require ideas far beyond   any weakly or strongly coupled quantum field theory used in our understanding of Nature. Given this, it seems natural to take a step back and question whether this may be the hint towards a more radical solution.

Abandoning  LI as a fundamental principle is a natural possibility in this direction.  Out of the many possible ways
of doing it,  we will  restrict to the case  where Lorentz symmetry is ``broken'' by the existence of a  preferred time direction at every point of space-time. The main motivation for this is that these theories contain the minimal ingredients to (hopefully) address the aforementioned issues of GR. Furthermore, they represent a well-defined physical framework  (existence of a preferred time direction) and have
an  interesting phenomenology.
Other possibilities include those studying preferred \emph{spatial} directions \cite{Ackerman:2007nb}, 
a direct breaking of the Poincar{\'e} group \cite{Cohen:2006ky}, or  the generic description inspired by the \emph{standard model extension} of particle physics \cite{Bailey:2006fd,Tasson:2014dfa}.

A particularly powerful way of defining the preferred time structure is to employ the formalism of \emph{spontaneously broken symmetries}. In this formalism, we introduce a vector field  $u^\mu$ with a non-zero time-like vacuum expectation value (vev)\footnote{We will not discuss the possibility of physics of unbroken phases (see e.g. \cite{Cheng:2006us} for related work in this direction). This is
consistent with the idea of dealing with an effective theory valid up to a cut-off energy. Even more, one can assume that only the broken phase exists even at high energies (Ho\v rava gravity).}.
 This vev is imposed by insisting that the vector possesses a fixed norm 
\be\label{eq:unit}
u_\m u^\m=1.
\ee 
The most concrete  proposal in this direction is \emph{Ho\v rava gravity} \cite{Horava:2009uw}. This framework assumes the existence of a preferred foliation of space-time into space \emph{and} time, whose normal 
vector provides the preferred time direction at each point.
On the other hand, a more generic method of  describing the low-energy phenomenology of gravitational theories with a preferred time direction
parametrised by $u^\m$ is to impose condition \eqref{eq:unit} via a constraint in the equation of motion itself (i.e. via a Lagrange multiplier).  This latter idea underlies the so-called \emph{Einstein-aether} theories introduced in Ref. \cite{Jacobson:2000xp}
(see also \cite{Gasperini:1987nq,Gasperini:1998eb} for related work).
One can relate the latter to the former by imposing the additional requirement that the Einstein-aether vector  $u^\m$ is normal to the hyper-surfaces of a constant field $\varphi$ (these hyper-surfaces define the preferred foliation), 
\be
\label{eq:kh}
u_\m\equiv\frac{\pd_\mu \varphi}{\sqrt{g^{\m\n}\pd_\m  \varphi\pd_\n \varphi}}.
\ee
The previous restriction yields the \emph{khronometric} theory \cite{Blas:2010hb}, which describes an independent family of 
effective theories valid at low-energies. 
   Hor\v ava gravity is understood in this framework 
as a UV completion associated to a restriction to a particular family of irrelevant operators \cite{Blas:2009yd}.

 Theories that violate Lorentz invariance (LV theories)  typically contain modes that propagate
faster than light (even instantaneously in Ho\v rava gravity \cite{Blas:2011ni}).
Recall that this is problematic in the Lorentz invariant case:  in flat-space,  this allows  to construct a closed time-like curve (CTC) using superluminal modes
and a series of boosts \cite{Tolman:Book}. However, this violation of causality is based both on superluminal propagation  \emph{and} on
the invariance of the theory under boosts.  If the latter is broken,  one 
can generate a  causal structure of space-time, thus preventing the formation of CTC, even in the presence of superluminal propagation. There is still a difference in the way this can happen in Einstein-aether or Ho\v rava gravity.
In Einstein-aether theory all the modes propagate with a finite speed \cite{Jacobson:2004ts}. Thus, the are light-cones defined at any point of space-time and
it is a dynamical question whether those allow for a well-defined causal structure. This is similar to the situation in GR, where not all solutions are free of CTC, and one expects the physical ones to be causal.  In Ho\v rava gravity, there are infinitely fast modes \cite{Blas:2011ni}. This is not a problem if one assumes that the  causal structure is fundamental and there is a preferred time direction in which the time evolution should be formulated. The well-posedness of the initial value problem is very different in this case, and  has not yet been clarified. For more details on causality and  the Cauchy problem in theories without Lorentz invariance, see \cite{Mattingly:2005re,Babichev:2007dw,Blas:2009yd,Donnelly:2010qd}.

The aim of this review is to  describe some of the consequences of the existence of the field $u^\m$   
in current experimental tests of gravitation. These can be divided into two domains: short or long distance modifications.
To define the boundary between the two, let us consider the generic theories including  $u^\m$ and a metric $g_{\m\n}$.  The
most relevant operators invariant under diffeomorphisms yields the action\footnote{We also assume CPT \cite{Blas:2010hb}. The condition \eqref{eq:unit} prevents from the existence of terms involving only $u^\m$ and $g_{\m\n}$ (without  derivatives). We assume that the only modification to GR comes from the presence of $u_\m$. Concerning  other possibilities, it seems worth mentioning that a formulation of a Higgs mechanism for massive gravitons is known only for theories including a preferred foliation (khronometric theories) \cite{Blas:2009qj}.} proposed in  \cite{Jacobson:2000xp}
\be 
\label{ae-action} S_{IR} = -\frac{M_0^2}{2}\int  ~d^{4}x \sqrt{-g}~ (R
+K^{\a\b}{}_{\m\n} \nabla_\a u^\m \nabla_\b u^\n+\lambda (u^\m u_\m-1))\,,
\ee 
where $g$ and $R$ are the metric determinant and the Ricci scalar and 
\be
K^{\a\b}{}_{\m\n} \equiv c_1 g^{\a\b}g_{\m\n}+c_2\delta^{\a}_{\m}\delta^{\b}_{\n}
+c_3 \delta^{\a}_{\n}\delta^{\b}_{\m}+c_4 u^\a u^\b g_{\m\n}\, ,
\ee 
($c_1$,
$c_2$, $c_3$ and $c_4$ being dimensionless coupling constants).  We emphasise the use of $M_0$ instead of $M_P$ for the mass scale
in front of the Einstein-Hilbert action to distinguish it from the quantity appearing in Newton's law -- see Section \ref{sec:solar},  in particular Eq.~\eqref{eq:Newton}. Finally, 
 we imposed the restriction to  the LV phase (\ref{eq:unit}) through a Lagrange multiplier $\lambda$. In the khronometric
case, this is not necessary since  (\ref{eq:unit}) is automatically satisfied (cf. \eqref{eq:kh}). Furthermore, the condition \eqref{eq:kh} implies that one of the terms in the previous action can be expressed in terms
of the others. It is customary to absorb the $c_1$ term into the other three terms, by multiplying the second, third and forth term respectively by the new couplings
\be\label{eq:EAtoKH}
\lambda\equiv c_2, \quad \beta\equiv c_3+c_1, \quad \alpha\equiv c_4+c_1.
\ee
 A particular version of Ho\v rava gravity 
that was popular in the past was the so-called \emph{projectable} case \cite{Horava:2009uw}. In our language
it corresponds to considering congruences with vanishing acceleration, which can be recovered from the general treatment with the 
parameters defined in \eqref{eq:EAtoKH} by taking  the limit $\alpha\to \infty$ \cite{Blas:2009qj}. We will not consider
this possibility in this review since it suffers from  formation of caustics and the presence of a very low
strong coupling scale which jeopardise its viability \cite{Blas:2009yd}.

A better  way to understand how to impose the  khronometric condition \eqref{eq:kh} in  the Einstein-aether case
was put forward in \cite{Jacobson:2013xta}. 
One can rewrite the aether part of the action \eqref{ae-action} in terms of more physically intuitive quantities related to the congruences of the field $u^\m$. For this, let us make the decomposition
\be
\nabla_\m u_\n=\frac{1}{3}\theta h_{\m\n}+\sigma_{\m\n}+\omega_{\m\n}+u_{\m}a_{\n},
\ee
where the \emph{spatial} metric is defined  as
\be
h_{\m\n}\equiv u_\m u_\n -g_{\m\n}.
\ee
The vector $a^\m\equiv u^\n \nabla_\n u^\m$ is the acceleration of the congruence, while $\theta\equiv \nabla^\m u_\m$ is the expansion. 
The shear $\sigma_{\m\n}$ and twist $\omega_{\m\n}$ are given  respectively by the symmetric and antisymmetric tensors
\be
\sigma_{\m\n}\equiv h^\a_{(\m} h^{\b}_{\n)}\left(\nabla_\a u_\b-\frac{1}{3}\theta h_{\a\b}\right),\quad 
\omega_{\m\n}\equiv h^\a_{[\m} h^{\b}_{\n]} \nabla_\a u_\b.
\ee
In terms of these quantities, one can write 
\be
\label{eq:geoaction}
\int  ~d^{4}x \sqrt{-g}~ (K^{\a\b}{}_{\m\n} \nabla_\a u^\m \nabla_\b u^\n)=\int ~d^4 x\sqrt{-g}\left(\frac{c_\theta}{3}\theta^2+c_\sigma \sigma^2+c_\omega \omega^2
+c_a a^2\right),
\ee
where
\be
\label{eq:cdef}
c_\theta\equiv c_1+c_3+3c_2, \quad c_\sigma\equiv c_1+c_3, \quad c_\omega\equiv c_1-c_3,\quad c_a\equiv c_1+c_4.
\ee
The interesting observation is that the twist term vanishes automatically if Eq.~\eqref{eq:kh} is satisfied. A way to impose this condition is to take the limit
\be
\label{eq:untwist}
c_\omega\to \infty,
\ee
with all other couplings held fixed. In this case, the dynamics of the system will always correspond to a twist-free vector field that can be written as
in Eq.~\eqref{eq:kh}.

Both for the Einstein-aether and khronometric cases, the cut-off scale at which the  
theory ceases to be described as a weakly coupled  quantum field theory  is  
\be
\label{MLV}
M_{LV}\equiv\sqrt{c_i}  M_0,
\ee
where we assumed that all the $c_i$ parameters are of the same order \cite{Blas:2010hb}.
If the LV extension addresses the problems of GR at short distances, it is natural to assume that $M_{LV}$  also 
represents the energy scale at which the dynamics of the GR degrees of freedom  are modified
 This means that all the 
predictions from GR presumably  change at energies above $M_{LV}$, and this 
is the scale we use to define the barrier between short or long distance LV modifications. As an example, in Ho\v rava gravity, at the scale $M_{LV}$ (or below)
the new operators related to the UV completion of the theory appear \cite{Blas:2009ck,Papazoglou:2009fj}. A characteristic operator would be 
\be
\label{eq:uvhorava}
S_{UV}=\frac{1}{M_*^4}\int d^4 x\sqrt{-g}\left[ (g^{\m\n}-u^\m u^\n)R_{\m\n}\right]^3,
\ee
with $M_*\leq M_{LV}$. Notice that $M_{LV}$ depends on the LV parameters. From different observations that will be described in Sec.~\ref{sec:large}, $c_i\ll 1$, which means that $M_{LV}\ll M_0$.

In the previous logic, the long distance modifications are  summarised by the expression \eqref{ae-action}. The field $u^\m$  possesses a vev and massless fluctuations around it which modify basically all possible gravitational experiments  (as we will see shortly).  This model has a very rich phenomenology and makes robust low energy predictions which can be experimentally falsified without the need to 
explore  the very high energies $M_{LV}$.
 
 Before moving to the body of the review, we would like to mention that the field  $u_\m$ can in principle be coupled 
to the particles in the standard model of particle physics (SM). This generates LV effects whose study has 
produced very strong bounds on the possible couplings \cite{Kostelecky:2008ts,Mattingly:2005re,Liberati:2013xla}.
The consequences of these bounds for gravitation are not universal.  If  LV affects identically both the SM and gravitation 
(including dark matter and dark energy) there is little hope to get bounds from the latter competitive with those of the SM. However, one can envisage
 mechanisms that separate the effects in these sectors, even  with respect to quantum corrections. This may happen 
 through a fine-tuning of parameters in the model but also due to (softly broken) supersymmetry  \cite{GrootNibbelink:2004za,Pujolas:2011sk}, 
 suppression mechanisms \cite{Pospelov:2010mp}
 or by a dynamical emergence of LI at low energies in the SM \cite{Bednik:2013nxa,Chkareuli:2001xe} (see \cite{Liberati:2013xla} for review).
We will adopt this second possibility in the  rest of the paper and assume that the fields in the SM do not interact directly with the field $u_\m$.

The structure of this work is the following: in Sec. \ref{sec:short} we review some consequences of the LV modifications
at energies larger than $M_{LV}$. The consequences for smaller energies are dealt with in Sec. \ref{sec:large}. We will start with 
a discussion of the degrees of freedom in Sec.~\ref{sec:dof}. Then we will describe how the dynamics in the Solar System are modified in Sec.~\ref{sec:solar}. In Sec.~\ref{sec:strong} we will describe the LV effects for compact objects, and emission of gravitation waves.
Finally, in Sec.~\ref{sec:cosmo} we will discuss cosmological implications. We will finish with some discussion and future directions.  

This review is aimed at giving an overview of the topic. For further details or to better understand what the constraints are we urge the reader to 
check the original literature that we cite.

A note about conventions: we will work with the $(+,-,-,-)$ signature of the metric. The speed of light will be set to $c=1$ except when explicitly displayed. We also chose units where $\hbar =1$.

\section{Modifications at short distances}\label{sec:short}

As we mentioned  in the previous section, we expect that the behaviour of gravity is completely modified at energy scales higher than $M_{LV}$. Since we will assume that the breaking of Lorentz invariance occurs within the  weakly coupled regime, one can start 
parameterizing the
changes in gravitation  by assuming that the linear equations for the perturbations of the metric  acquire LV terms. Assuming that parity and time inversion are not violated and that the linear equations
are  at most second order in time derivatives (to ensure  that no Ostrogradski-like instability is present), we can summarise these effects by using  the dispersion relations
\be
\label{eq:lvgw}
\omega^2=p^2\left(1+\sum_{n=1}^{L}\alpha_n\left(\frac{p}{M^{gw}_{LV}}\right)^{2n}\right),
\ee
for the propagating degrees of freedom (e.g the graviton) and the modified Poisson's equation 
\be
\label{eq:new}
p^2\left(1+\sum_{n=1}^{L}\beta_n\left(\frac{p}{M^{\phi}_{LV}}\right)^{2n}\right)\phi=-\frac{1}{2M^2_0}\tau_{00},
\ee
for the potentials $\phi$  sourced by matter's energy, represented by $\tau_{00}$. We will assume that the dimensionless LV parameters $\alpha_i$ and $\beta_i$ in the previous expressions are all ${\cal O}(1)$. The vanishing of these parameters return the theory to the usual LI one. We have made the difference between $M^{\phi}_{LV}$  and $M^{gw}_{LV}$
to emphasise that the two quantities may be independent (also the constraints on them may come from different observations).

Let us first discuss the modifications of the graviton's dispersion relation, Eq.~\eqref{eq:lvgw}. Though 
 no competitive bounds have been produced yet, these can arise in the future from 
 different observations.  First, if the gravitational waves (GW) have  the dispersion relation \eqref{eq:lvgw}, this modifies the frequency dependence in the 
 propagation of the wave-fronts, which may be detected by future detectors of GW   \cite{Mirshekari:2011yq}. 
  Furthermore, any process emitting GW with energies $E\gtrsim M^{gw}_{LV}$ should also be altered if these have the modified dispersion relation we are studying. However, if we assume that $M^{gw}_{LV}\approx M^{\phi}_{LV}$, 
and given that the latter are constrained to be  $M^{\phi}_{LV}< (\mu m)^{-1}\approx 10^{-2} \ \mathrm{eV}$  (see below), 
 GWs produced by astrophysical sources never reach  such high energies\footnote{ We remind the reader
 that this is the energy of the gravitational modes. The relation to the typical 
 energies in the matter sector of the sources comes from Einstein's equation, and is thus suppressed by powers of $M_0$ \cite{Yagi:2013ava}.} \cite{Yagi:2013ava}.

This is different  for the GWs generated in the primordial universe. In this case, the typical energies during production may be almost as high as $M_P$. Thus, if  primordial GWs are observed, the range of energies  at which LV has been tested  (in fact any short-distance modification) would improve dramatically. 
 One can easily show that
 LV primordial GWs (or a LV inflaton with dispersion relation~\eqref{eq:lvgw}) would have a similar spectrum than for the standard case except at energies very close to $M_{LV}^{gw}$
 (see e.g \cite{Lim:2004js,Mukohyama:2010xz,Wang:2010an,Brandenberger:2012aj,Barrow:2014fsa} for further discussion on modified dispersion relations in cosmology). Another observable sensitive to deviations with respect to GR is the bispectrum of GWs and its consistency conditions derived in standard inflationary models (and their possible violations)  \cite{Adshead:2009bz,Maldacena:2002vr}. It would be extremely interesting to understand how the bispectrum of GWs is modified in LV theories,  which 
 for Gaussian initial conditions requires an analysis  beyond the dispersion relation \eqref{eq:lvgw} and including the leading non-linearities of the theory.

Much more is known about possible deviations of the potentials at high-energies, Eq.~\eqref{eq:new}. To get a better handle in the way these deviations
translate into modifications to Newton's law, let us focus on the case  relevant for Ho\v rava gravity where only  $\beta_2\neq 0$, and absorb its value into $M_{LV}^\phi$ ($M_{LV}^\phi\mapsto  M_{LV}^\phi \beta_4^{1/4}2^{1/8}$) to write
\be
p^2\left(1+\left(\frac{ p}{\sqrt{2}M^{\phi}_{LV}}\right)^{4}\right)\phi=-\frac{1}{2M^2_0}\tau_{00}.
\ee
Other situations in the context of \cite{Bailey:2006fd} have been recently discussed in \cite{Bailey:2014bta}.
Taking a point particle at rest as a source,  $\tau_{00}(x,t)=m_{pp}\delta^{(3)}(x^i)$, we find that away from the source 
\be
\label{eq:short}
\phi=-\frac{m_{pp}}{8\pi M_0^2 R}\left[1-e^{-M_{LV}^\phi R}\cos\left(M_{LV}^\phi R\right)\right],
\ee
where $R$ is the distance from the source. 
This potential is very interesting because it regularises the divergent behaviour of Newton's potential at small distances. Furthermore,
at scales where the deviations start to be important,  it is similar
to the potentials that have been considered  to constrain the deviations from Newton's law at short distances \cite{Kapner:2006si,Murata:2014nra,Will:2014xja}
\be
\label{eq:shortpot}
\phi=-\frac{m_{pp}}{8\pi M_0^2 R}\left[1+\tilde\alpha\, e^{-R/\tilde \lambda}\right].
\ee
\begin{figure}
\begin{center}
\includegraphics[width=0.5\textwidth]{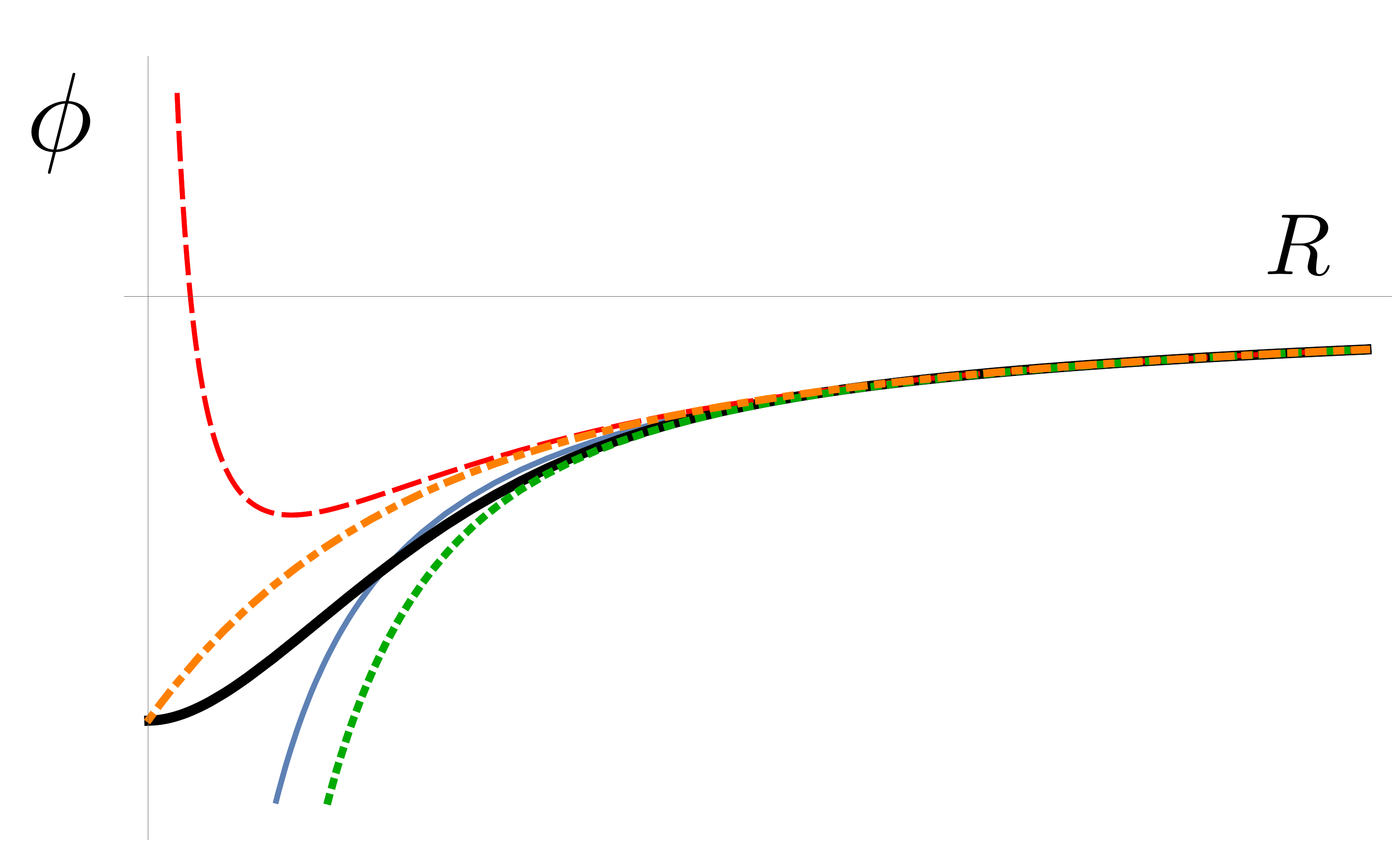}
\caption{\label{fig1}
Shapes of the potentials for the modifications of Newton's law discussed in the text. The thin solid blue line represents the standard Newtonian potential behaving  as $1/R$. The thick solid black line shows the form \eqref{eq:short} related to the LV modifications. This solution is not singular at the origin. The other  lines represent three cases of the parametrised form \eqref{eq:shortpot} with 
$\tilde \lambda=1/M_{LV}^{\phi}$, and $\tilde \alpha<-1$ (red, dashed), $\tilde \alpha>-1$ (green, dotted)  and $\tilde \alpha=-1$ (orange, dot-dashed).} 
\end{center}
\end{figure}
From these works, one concludes that a bound of the form $M_{LV}^\phi \lesssim (\mu m)^{-1}$ should apply.  
As shown in Fig.~\ref{fig1}, the differences of the form assumed to get this bound with the potential \eqref{eq:short} are important (e.g. the potentials  in \eqref{eq:shortpot} are singular at short distances except for $\tilde\alpha=-1$), and it would be interesting to reprocess the experimental data to get the precise bound on $M_{LV}^\phi$. As shown in Fig.~\ref{fig1}, one can see that the oscillations related to the cosine in \eqref{eq:short} are invisible at large distances.

Similar to the GW case, one also expects  any phenomena exciting very energetic  wave-modes of the gravitational field (in this case of the potentials) to  be sensitive to the modifications in  Eqs.~\eqref{eq:new}.  Natural candidates are  the primordial universe and very compact objects. For the
former, we do not know of any dedicated discussion. For the latter, the bounds coming from table-top experiments on $M_{LV}^\phi$ makes any phenomenological observation in these systems challenging  \cite{Yagi:2013ava} (cf. the discussion above). 

Finally, even though our previous discussion was organised around the modified equations \eqref{eq:lvgw} and 
\eqref{eq:new},  the effects of LV at short distances (high energies)  may also be important 
for the  background evolution in the primordial universe.  
These  may be relevant to resolve cosmological singularities  or to generate bouncing solutions.   
For the case  of Ho\v rava gravity, we refer the reader to the works 
\cite{Calcagni:2009ar,Kiritsis:2009sh,Mukohyama:2010xz,Brandenberger:2009yt} for further detail (in  \eqref{eq:uvhorava} we wrote a possible new contributions to the Lagrangian). Similarly, the 
 background evolution during inflation may be modified by LV effects at high energies. This may come from the modifications of gravity coming from new terms (cf. \eqref{eq:uvhorava}) or from the possible coupling of the inflaton to the field $u^\m$.  We are not aware of works discussing the first possibility. 
 Concerning possible couplings of
 the aether to the inflaton, they have been considered only in the cases where the aether action is of the form \eqref{ae-action}, and thus we will comment on this in the next section (Sec.~\ref{sec:inflation}).

\section{Modifications at large distances}\label{sec:large}

As we explained before, LV theories of gravity deviate from GR also at energy scales smaller than $M_{LV}$. This is expected  since Lorentz invariance in GR is associated to the group of diffeomorphisms. In LV theories this gauge invariance is broken to a smaller group, which translates into the presence of new degrees of freedom. Furthermore, around Minkowski space, the boosts of the global Lorentz group are broken  ``spontaneously'' which suggests that the new modes will be massless Goldstone bosons\footnote{Recall that since we are dealing with LV theories, 
the counting and properties of such fields are different from  
the Lorentz invariant case, see e.g.  
\cite{Brauner:2010wm,Nicolis:2012vf,Watanabe:2012hr}.}. As will become clear shortly, independently of these fluctuations the existence of the vev of $u^\m$ will have phenomenological implications  (e.g.  it generates violations of the  equivalence principle). In this section we will focus on the action \eqref{ae-action} and extract its consequences for different tests of gravity.

\subsection{Degrees of freedom and stability constraints}\label{sec:dof}

Out of the different constraints, we can first single out those related to the stability of different configurations. 
One of the basic requirements to assess the viability of the theory is  the existence and stability of the backgrounds
observed in cosmological and local physics. This translates into imposing that Minkowski space or de Sitter space with 
curvature given to the cosmological constant should be stable solutions of the theory. In this section we will focus on Minkowski space.

One can easily check that  $g_{\m\n}=\eta_{\m\n}$ and $u^\m=\delta^\m_0$ is in fact a solution of the equations of motion derived from  the action \eqref{ae-action}. The perturbation spectrum around this solution consists of not only the original tensor graviton modes, but also propagating vectors and a scalar modes characterised by the velocities  \cite{Jacobson:2004ts} (see
\eqref{eq:cdef} for the  connection to the parameters in \eqref{ae-action}), 
\be
\label{eq:speeds}
 \varv_2^2= \frac{1}{1-c_\sigma}, \quad \varv_1^2=\frac{c_\sigma+c_\omega-c_\sigma c_\omega}{2c_\sigma(1-c_\sigma)}, \quad \varv_0^2=\frac{(c_\theta+2c_\sigma)(1-c_a/2)}{3c_a(1-c_\sigma)(1+c_\theta/2)}.
\ee
Notice that the graviton's velocity is modified. These velocities should all be real so that there are no tachyon-like instabilities. To avoid Cherenkov radiation requires that they are close to or greater than the speed of light \cite{Elliott:2005va} -- otherwise the ultra-high energy cosmic rays would lose energy
by emission of the subluminal modes\footnote{ The precise calculation puts a bound on how subluminal the 
signal can be, given the  observed energies in cosmic rays. This bound  depends on the coupling of the SM fields to the different modes through gravity. Since the result is very close to imposing luminal or superluminal propagation, we prefer to directly work with the latter assumption and refer the interested reader to 
  \cite{Elliott:2005va} for further details.}.
We can simplify things considerably by restricting to the subfamily of parameters  compatible with Solar System tests (to be discussed in Sec.~\ref{sec:solar}). This corresponds to the conditions 
\be
\label{eq:solar}
c_a=2\frac{c_\omega c_\sigma}{c_\sigma+c_\omega}, \quad c_\theta=-c_a.
\ee
which should be valid up to deviations of order ${\mathcal O}(10^{-4})$.
Then the stability conditions  at this order reduce to
\be
\label{stab1}
0\leq c_\s\leq 1, \quad 0\leq c_\omega \leq \frac{c_\s}{3(1-c_\s)}.
\ee
One  must also impose the condition 
\begin{equation}
\label{stab2}
c_{\sigma}+c_{\omega}>0
\end{equation}
such that the vector and scalar modes have positive energy \cite{Lim:2004js,Eling:2005zq}.

Further constraints  could be derived by requiring  that certain physical configurations have positive energy. In particular, it was shown in
\cite{Garfinkle:2011iw} that the previous constrains are enough to ensure the positivity of energy for
 solutions where $u^\m$ is hypersurface-orthogonal and where one of those hyper-surfaces is asymptotically flat and has vanishing trace of
the extrinsic curvature, which corresponds to $\theta=0$ in the language of \eqref{eq:geoaction}.

\subsection{Post-Newtonian formalism and Solar System constraints}\label{sec:solar}

The first system of interest to test gravitation is  within the Solar System, where  there are many different observations at disposal \cite{Will:2014xja}. For this, the non-relativistic approximation $v\ll c$ (with 
$v$ being a typical velocity of the system and $c$ the speed of light) and small gravitational fields is extremely good. One also uses the fact that for  virialized systems  the strength of the
gravitational field is related to $v/c$.
In this case, one can organise the calculation of gravitational potentials in the so called post-Newtonian (PN)
expansion \cite{Blanchet:2006zz,Will:Book,Wein:Book}. 
In order to solve the equations completely in the PN framework for the action \eqref{ae-action} we need to specify the velocity of the sources (the Sun in this case) with respect to the preferred frame specified by $u^\mu$ (in particular its PN order). The most common (and natural) candidate is to assume that the preferred frame is aligned with the cosmic microwave background (CMB) frame \cite{Will:2014xja}. This automatically follows from assuming isotropy,  which selects $u^\mu$ to be
 the hypersurface orthogonal with respect to the FRW spatial foliation. Remarkably, this configuration is an attractor solution of the evolution in case of slight misalignment \cite{Kanno:2006ty,Carruthers:2010ii}.
Under these conditions, the metric at PN order reads \cite{Foster:2005dk,Blas:2011zd}
\be \label{eq:PPNmetric}
\begin{split}
g^{PPN}_{00}&=1-2U+2\beta^{PPN}U^2-4\Phi_1-4\Phi_2 -2\Phi_3-6\Phi_4,\\
g^{PPN}_{ij}&=-(1+2\gamma^{PPN}U)\delta_{ij}, \\
g^{PPN}_{0i}&=\frac{1}{2}(7+\alpha_1^{PPN}-\alpha_2^{PPN})V_i^{PPN} +\frac{1}{2}(1+\alpha_2^{PPN})W_i^{PPN},\\
\end{split}
\ee
where the relevant potentials are  (the rest of potentials can be found in \cite{Will:Book})
\begin{gather}   \nonumber
  U(x) =G_{\rm N} \int d^3y \frac{\rho(y)}{|x-y|},\\
  V_i^{PPN}=G_{\rm N} \int d^3y \frac{\rho(y)v^i}{|x-y|}, 
  \quad W_i ^{PPN}=G_{\rm N} \int d^3y \frac{\rho(y)}{|x-y|}\frac{v_j (x_j - y_j)(x^i -y^i)}{|x-y|^2},
  \label{eq:PPNv}
\end{gather}
where $\rho(r)$ is the energy density of the source and $G_{\rm N}$  is defined by
\be
\label{eq:Newton}
G_N\equiv \frac{1}{4\pi M_0^2 \left(2-c_a\right)}.
\ee
The different coefficients in \eqref{eq:PPNmetric} are part of the parameterized-post-Newtonian (PPN) formalism \cite{Will:Book}. For the Einstein-aether case these read  \cite{Foster:2005dk}
\beqa
&\gamma^{PPN}=\beta^{PPN}=1, \\ \label{alpha1}
&\alpha_1^{PPN}=\alpha^{\AE}_1 \equiv4\left(\frac{c_\omega(c_a-2c_\sigma)+c_a c_\s}{c_\omega(c_\s
-1)-c_\s}\right)\,, \quad \alpha_2^{PPN}= \alpha^{\AE}_2 \equiv \frac{\alpha^{\AE}_1}{2} +\frac{3(c_a-2 c_\sigma)(c_\theta+c_a)}{(2-c_a)(c_\theta+2c_\sigma)}\,.
\eeqa
To obtain the equivalent expression for the khronometric theory \cite{Blas:2011zd}, it is enough to take the limit \eqref{eq:untwist}. The  GR limit corresponds to $\gamma^{PPN}=\beta^{PPN}=1$ and $\alpha_1^{PPN}=\alpha_2^{PPN}=0$. Recall that in the   PPN formalism, the presence of non-vanishing $\alpha_1^{PPN}$ and $\alpha_2^{PPN}$ is the trademark at PN level of LV for any
 semi-conservative theory (e.g. theories with a notion of conserved energy) \cite{Will:Book}.

As we stressed, the form of the metric \eqref{eq:PPNmetric} can be used to describe gravity in the  Solar System.  The different test of gravitation yield constraints
for the deviations of GR of the form $|\alpha^{PPN}_1|\lesssim 10^{-4}$ and
$|\alpha^{PPN}_2| \lesssim 10^{-7}$, coming from lunar laser ranging
and solar alignment with the ecliptic ~\cite{Will:2014xja,Will:2014bqa}. Those are very small numbers as compared with the rest of bounds to be discussed in this review, which  is why we will often work in the region of parameter space where they are identically zero.
This corresponds to the  conditions considered in \eqref{eq:solar}.

A subtlety occurs when \eqref{eq:solar} are imposed in the khronometric case. In fact, from \eqref{alpha1} it is easy to see that in the limit $c_\omega\to \infty$ both $\alpha_2^{\AE}$ and
$\alpha_1^{\AE}$ are proportional to the combination $c_a-2c_\sigma$. Thus, it is enough that this combination cancels to satisfy the Solar System constraints. Something similar happens when one wants to saturate these constraints. For illustration, let us do it in the limit where all the parameters are order one or smaller: first
one  imposes $\left|\alpha^{\AE}_1\right| < 10^{-4}$, which corresponds to $|c_a-2c_\sigma|< 10^{-4}$. The other condition to be satisfied is $\left|\alpha^{PPN}_2/\alpha^{PPN}_1\right|< 10^{-3}$, which one can do by making the previous constraint stronger, $|c_a-2c_\sigma|< 10^{-7}$, or by imposing the extra condition on $c_\theta+c_a$ (cf. \eqref{alpha1}). In the first case a single condition is enough to satisfy the Solar System tests, which leaves an unconstrained two-parameter space.

\subsection{Strongly gravitating objects }\label{sec:strong}

\subsubsection{Conservative dynamics}

Lorentz violating effects are also important for the dynamics of compact objects.  
If we are only interested in the effects far away from the sources, an efficient
way to describe them is by noticing that the coupling between the field
$u^\m$ and the gravitational fields will affect the dynamics of the full body. Thus,  far away from the sources, where they can be considered as point-like,  this coupling can be described as an
effective interaction of the source with the aether field \cite{Foster:2007gr,Yagi:2013ava} (see
\cite{Damour:1992we} for similar ideas in  scalar-tensor theories). In this approximation the action for the source can be described as a point-particle with an aether charge 
\be
\label{eq:acsens}
S_{\rm pp \; A}=-\int ds_A \tilde m_A(\gamma_A), 
\ee
where $\tilde m_A$ is a function of the Lorentz factor of the source with respect to the aether, $\gamma_A\equiv u_\m v_A^\m$, with $v_A^\m$ being the four-velocity of the source. Finally, $ds_A$ is the line element of the trajectory. One can make further progress by assuming that $\gamma_A$ is close to one (which corresponds to a slight misalignment of the two 4-vectors), 
\begin{align}
\label{taylor-pp}
S_{\rm pp \; A} &=- \tilde{m}_{A}  \int ds_A \; 
\left\{1 + \sigma_{A} \left(1- \gamma_{A}\right)
+ {\cal{O}}\left[\left(1-\gamma_{A}\right)^{2}\right] \right\}\,, 
\end{align}
where we have defined the sensitivity parameters $\sigma_{A}$ via %
\begin{align}
\label{sigma-def}
\sigma_{A} &\equiv - \left.\frac{d \ln \tilde{m}_{A}(\gamma_A)}{d \ln \gamma_{A}}
\right|_{\gamma_{A} = 1}\,.
\end{align}
The previous action  can be used to describe the dynamics of different systems. For instance, for a binary system with point sources $A$ and $B$ with positions $x^i_A$  and $x^i_B$, at first Newtonian order one finds that the velocity of the object $A$, $v^i_A\equiv \dot x^i_A$, satisfies (a similar expression holds for $\dot{v}_B^i$ after exchanging $A \leftrightarrow B$)~\cite{Foster:2007gr}
\begin{equation}
\label{eq:New_active}
\dot{v}_A^i= -\frac{{\cal G} {m}_B  (x^i_A-x^i_B)}{|\boldsymbol{x}_A-\boldsymbol{x}_B|^3}\,,
\end{equation}
where   we defined the {\textit{active}} gravitational masses as
\be
\label{active-mass}
m_B\equiv\tilde{m}_B (1+\sigma_B),
\ee
and the 2-body coupling constant
\be
\label{calG}
{\cal G}\equiv\frac{G_N}{(1+\sigma_A) (1+\sigma_B)}\,.
\ee
One sees from these expressions how the strong equivalence principle is violated in this system ($m_A\neq \tilde m_A$ as a consequence of $\sigma_A\neq 0$).

 The action \eqref{taylor-pp} also allows for the discussion of
PN effects in objects with strong gravitational fields. For this, we consider an isolated object and
compute its gravitational field  far away from
it, where it can be considered as a point.  At the PN level, our metric will be of the form \eqref{eq:PPNmetric} \cite{Will:Book}, 
but this time, and in contrast to GR, the PPN parameters will depend on the internal structure of the body through the sensitivities $\sigma_A$
\cite{Damour:1992we,Damour:1992ah}. This is a consequence of the violation of the \emph{strong} equivalence principle.  
One can show that the new PPN parameters  $\alpha_i^{PPN}=\hat \alpha_i^{\AE}$ read  \cite{Foster:2007gr,Yagi:2013ava}
\begin{align}
\label{SF-alpha1}
\hat{\alpha}_{1}^{\AE} &= \alpha_1^{\AE} + \frac{c_\omega (8+\alpha_1^{\AE}) \sigma_A^{\AE}}{2 c_1}\,,
\\ 
\label{SF-alpha2}
\hat{\alpha}_{2}^{\AE} &= \alpha_2^{\AE} + \frac{c_\omega (8 + \alpha_1^{\AE} ) \sigma_A^{\AE}}{4 c_{1}} - \frac{  (c_a -2)  (\alpha_1^{\AE} - 2 \alpha_2^{\AE}) \sigma_A^{\AE}}{2  (c_a - 2 c_\sigma)}\,.
\end{align}

 These quantities can be constrained very efficiently with data from binary and solitary pulsars \cite{Damour:1992ah,Weisberg:2013pha,Shao:2013wga,Shao:2012eg}. By using the orbital dynamics of the pulsar-white
dwarf binary PSR J1738+0333 \cite{Freire:2012mg}, one  gets $\left|\hat \alpha_1^{PPN}\right|< 10^{-5}\quad (95\%\ \mathrm{CL})$ \cite{Shao:2012eg}. Using data from solitary pulsars one can get the bounds 
$\left|\hat \alpha_2^{PPN}\right|< 1.6\cdot 10^{-9}\quad (95\%\ \mathrm{CL})$ \cite{Shao:2013wga}. These bounds are very strong, but to translate them into constraints for the fundamental parameters of \eqref{ae-action}, one needs to solve for the sensitivities (cf. \eqref{SF-alpha1} and \eqref{SF-alpha2}).

This requires solving the field equations for  realistic sources and matching the metric to the form \eqref{eq:PPNmetric} in the region where the PN expansion applies\footnote{Recall that we are including the strong non-Newtonian effects in the sensitivities. The rest of corrections follow the PN logic.}. 
Then 
one simply uses the expressions \eqref{SF-alpha1} and \eqref{SF-alpha2} to compute the sensitivities.
The first results in this direction appeared in \cite{Foster:2007gr}. This was an analytical calculation based on the weak-field limit, i.e.~expanding
in the ratio of the binding energy $\Omega$,
\be 
\Omega \equiv - \frac{1}{2}  G_N \int d^3x \rho (r) \int d^3x' \frac{\rho(r')}{|x-x'|}\,,
\ee
 to the mass of the object $\tilde m$. One can show
that the sensitivity scales as
\be
\label{eq:weak-field-AE-s}
s^{\AE}_A = \left( \alpha_1^{\AE} - \frac{2}{3} \alpha_2^{\AE} \right)
\frac{\Omega_A}{\tilde m_A} + \mathcal{O} \left( \frac{\Omega_A^2}{\tilde m_A^2},
\right)\,, \ee
where we have introduced  the function 
\be
s_A\equiv \frac{\sigma_A}{1+\sigma_A}\,.
\ee

\begin{figure}
\begin{center}
\includegraphics[width=1\textwidth]{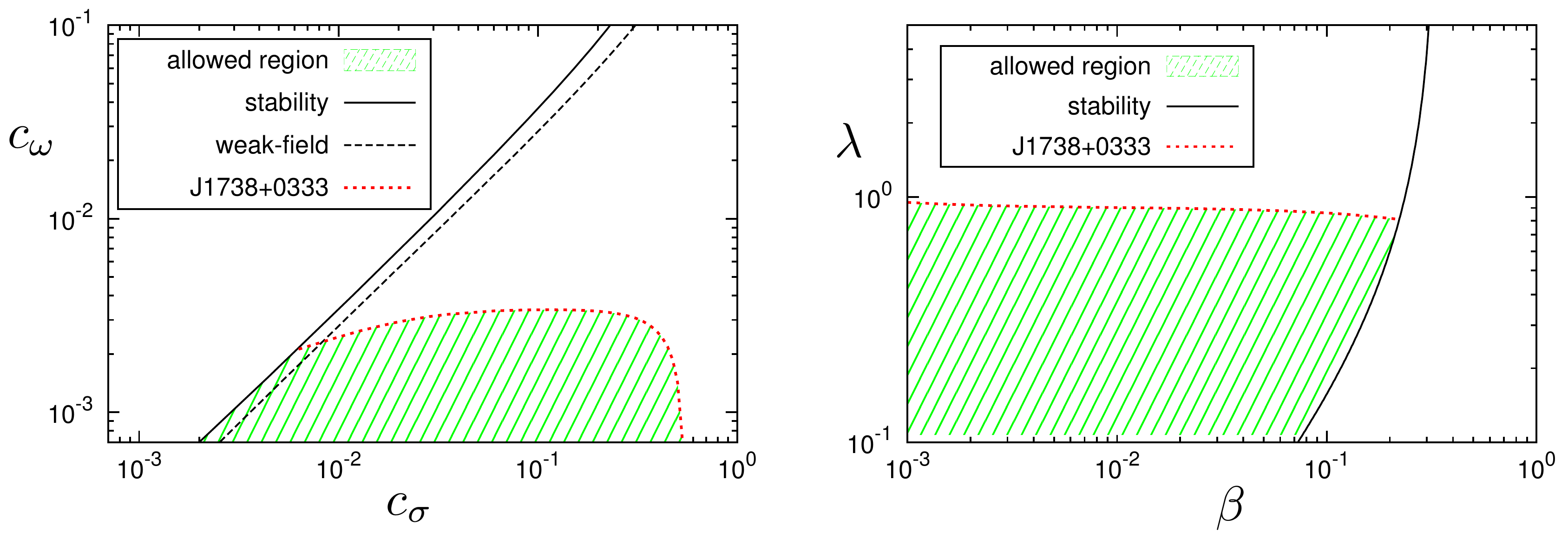}
\caption{\label{fig2}
Constraints on the LV parameters coming from the pulsar-white
dwarf binary PSR J1738+0333 \cite{Freire:2012mg}. The left panel corresponds to the Einstein-aether case, while the right one is the khronometric one. The green-shaded regions 
corresponds to the allowed regions after using the data from PSR J1738+0333 and once the Solar System  \eqref{eq:solar}, Cherenkov, and stability constraints \eqref{stab1} 
and \eqref{stab2} have been imposed. For comparison, the allowed regions after imposing the Cherenkov, stability and Solar System constraints are
those to the right of the solid  black line in the Einstein-aether case and to the left of the solid black line in the khronometric case. In the
left panel, the dashed-line corresponds to the combination of parameters where the orbital decay (see Sec.~\ref{sec:emission}) agrees exactly with that of GR. The figures are taken from \cite{Yagi:2013ava}.
}
\end{center}
\end{figure}
This approximation is   not always  applicable in realistic situations with 
very compact objects (big $\Omega$). 
In Refs.  \cite{Yagi:2013qpa,Yagi:2013ava}, the metric outside  spherically-symmetric, non-rotating, cold (and thus old) neutron stars was computed for different equations of state\footnote{See also \cite{Eling:2007xh} for the 
 first work studying neutron stars in Einstein-aether theories.}. It was also  assumed a small Lorentz factor $\gamma_A$ with respect to the preferred frame. This is required for one  
to remain in the same approximation used to define the sensitivities in \eqref{taylor-pp}, and in which the expressions \eqref{SF-alpha1} and \eqref{SF-alpha2} were derived.   At zeroth order in the velocities one gets modifications of the profile of the star (coming from  the differences in  the Tolman-Oppenheimer-Volkoff equations, see e.g. \cite{Wein:Book}).  However, these modifications are not enough to produce strong constraints, cf. \cite{Yagi:2013ava}.
The next corrections appear at first order in the velocity of the star with respect to the preferred frame. To remain in the PN
approximation, one assumes
this velocity to be small with respect to $c$. In this case, by comparing the numerical solution with the asymptotic metric 
  \eqref{eq:PPNmetric} with parameters  $\alpha_i^{PPN}=\hat \alpha_i^{\AE}$, one can read the value of $\sigma_A$ after using  \eqref{SF-alpha1} and \eqref{SF-alpha2} \cite{Yagi:2013qpa,Yagi:2013ava}.
 The objects for which the previous approximation holds are well represented  within the current data set of pulsars (solitary or in binaries). However, to extract the sensitivities one needs to  know the mass, even if the equation of state is known. The latter can be extracted in binary systems, but not in isolated pulsars.
 Quite interestingly, the dynamics of the  pulsar-white
dwarf binaries PSR J1738+0333 \cite{Freire:2012mg} already put very strong constraints in the LV parameters, breaking the degeneracies remaining after imposing the
 Solar System bounds \eqref{eq:solar}. The constraints  found in \cite{Yagi:2013ava} for this case are shown in Fig.~\ref{fig2}.

 Finally, before closing this section, we would like to mention that further phenomenological input in the theory may come from 
 the differences in black hole solutions \cite{Barausse:2011pu} (see also \cite{Garfinkle:2007bk,Barausse:2013nwa}). In fact, even if the deviations
away from the Schwarzschild metric are typically no more than a few percent for most of the allowed 
 parameter regions (which makes them difficult to be observed with electromagnetic probes, such 
 as such as accretion disk spectra, iron-K$\alpha$ lines or gravitational lensing  \cite{Barausse:2011pu})
 these may be within reach of future gravitational-wave detectors 
from the observations of 
extreme mass-ratio inspirals \cite{Barausse:2011pu}.
 
\subsubsection{Dissipative effects: emission of gravitational waves}\label{sec:emission}
 
Another  key prediction of GR that has been tested with increasing precision in the last thirty years is the damping of orbits in binary systems coming from the emission of gravitational radiation \cite{Will:2014xja,Will:2014bqa}. The best known and studied of these systems is  the Hulse-Taylor binary pulsar
PSR 1913+16  \cite{Hulse:1974eb,Taylor:1989sw}. For these systems, not only the conservative dynamics are modified in LV theories (as described in the previous section), but also the emission of gravitational waves. This arises not only from the modifications in the    degrees for freedom already present in GR (e.g. the speed of the tensor modes is given by $\varv_2$ in  \eqref{eq:speeds}), but also from the existence of new light degrees of freedom  in the theory. 

The first calculations of the power emitted by binary systems in LV theories of gravity were performed in \cite{Foster:2006az,Blas:2011zd}.
However, these works focused on  Newtonian sources for which the gravitational radiation is very weak, and its effects in the damping of the orbits of the systems have not been observed yet. To get sizable effects, one needs
to consider  binary systems with two compact objects (systems with two pulsars or one pulsar and one white-dwarf are ideal). This program was undertaken in \cite{Foster:2007gr}, except for the calculation of the sensitivities beyond the limit \eqref{eq:weak-field-AE-s} (and barred some typos corrected in \cite{Yagi:2013ava}). 
For an orbit of two bodies $A$ and $B$ with small eccentricity, with relative velocity $v$ and center of mass velocity $V_{CM}$ with respect to the preferred frame, at first PN
 order one finds
that the  rate of change in the orbital period of the system $P_b$ due to the emission of gravitational waves is
\be
\frac{\dot{P}_{b}}{P_{b}} = - \frac{192 \pi}{5} \left(\frac{2 \pi G_{N} m}{P_{b}}\right)^{5/3} \left(\frac{m_A m_B}{m^2}\right) \frac{1}{P_{b}} \left<\mathcal{A}\right>\,,
\label{flux-AE}
\ee
where $m_{A}$ are the active masses \eqref{active-mass}, $m = m_{A} + m_{B}$ is the total active mass, and we have defined
\begin{align}
\label{A-AE}
\left<\mathcal{A}\right> \equiv &
\frac{5(1-c_a/2)}{32} \left(s_A- s_B\right)^2  \left( \frac{P_b}{2 \pi G_N m} \right)^{2/3} \mathcal{C} 
 \times \left[ 1 + {\cal{O}} \left( \frac{v^{2}}{c^{2}}, \frac{V_{CM}^{2}}{c^{2}},(s_{A}-s_{B})^{-1} \frac{V_{CM} v}{c^{2}} \right) \right]
\nn \\
&+  \left( 1-\frac{c_a}{2} \right)\left[\left(1 - s_{A}\right) \left(1 - s_{B}\right)\right]^{2/3}  
\times
\left( \mathcal{A}_1 + \mathcal{S} \mathcal{A}_2 + \mathcal{S}^{2} \mathcal{A}_3  \right) \left[ 1 + {\cal{O}} \left( \frac{v^{2}}{c^{2}} \right) \right]\,.
\end{align}
The coefficients used above are
\begin{align}
\label{A-funcs-1}
{\cal A}_1 &\equiv \frac{1}{\varv_2}+\frac{2 c_a c_\sigma^2}{(2c_1-c_\omega c_\sigma)^2 \varv_1}+\frac{3 c_a(Z-1)^2}{2 \varv_0(2-c_a)}\,,  \  \ {\cal A}_2 \equiv \frac{2(Z-1)}{(c_a-2)\varv_0^3}+\frac{2 c_\sigma}{(2c_1-c_\sigma c_\omega)\varv_1^3}\,,  
\\ 
\label{A-funcs-3}
{\cal A}_3 &\equiv\frac{1}{2 \varv_1^5 c_a}+ \frac{2}{3c_a(2-c_a)\varv_0^5}, \ \ {\cal C} \equiv \frac{4}{3 \varv_0^3
  c_a(2-c_a)}+\frac{4}{3 c_a \varv_1^3},
  \end{align}
  with  
  \be 
\label{Z-def}
Z \equiv \frac{(\alpha^{\AE}_1-2\alpha^{\AE}_2)(1-c_\sigma)}{3(2c_\sigma-c_a)}, \quad {\cal{S}} \equiv s_A m_B/m +s_B m_A/m\,.
\ee
GR is recover when ${\cal A}_2={\cal A}_3={\cal C}=0$, ${\cal A}_1=1$ and the LV parameters and sensitivities vanish. From \eqref{flux-AE} one can already 
understand why it is advantageous to stay in the class of binaries with small eccentricities: 
for these systems the dependence on the unknown  $V_{CM}$  is suppressed and the constraints are independent of it (provided that $V_{CM} \lesssim v$, and  that we stay within  the PN approximation). Furthermore, 
this approximation is also interesting because the most relativistic binary pulsar observed, the
double binary pulsar PSR J0737-3039 \cite{Kramer:2006nb}, has close to zero eccentricity. Finally, the inclusion 
of eccentricity in the analysis would require to calculate not only the effect in the orbits of the momentum flux put also that of 
the angular momentum flux. This makes the analysis much more involved\footnote{We thank E. Barausse and N. Yunes for discussions on this point}.

The first important thing to notice about the expression \eqref{A-AE} is that the term proportional to $\cal C$ corresponds to a dipolar contribution that is enhanced with respect to the quadrupolar term by ${\cal O}(c/v)^2$, which translates into the factor $\left( \frac{P_b}{2 \pi G_N m} \right)^{2/3}$. The fact that this contribution is multiplied by the difference of the sensitivities squared means that it will be important only for asymmetric systems (e.g. a binary of a neutron star and a white dwarf).  These differences (in fact the parameters $s_i$ themselves) are typically very small which is why we did not consider the next
PN corrections to the dipole term in \eqref{A-AE}.
 Notice also how the quadrupolar contribution is modified by both the change in the speed of the graviton being $\varv_2$ and the presence of other degrees of freedom (contributions depending on $\varv_1$ and $\varv_0$).
\begin{figure}
\begin{center}
\includegraphics[width=.8\textwidth]{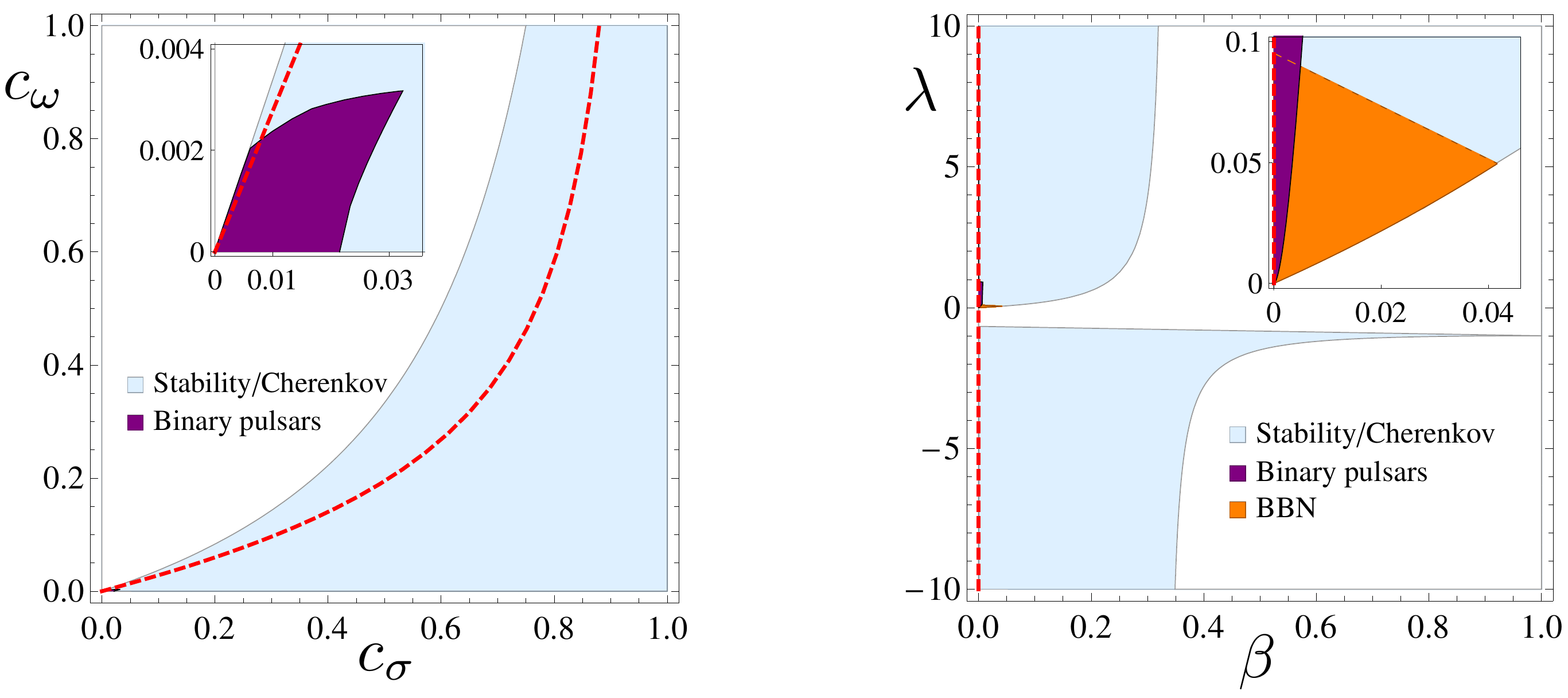}
\caption{\label{fig3}
Constraints on the LV parameters coming from pulsars (both solitary and binaries). The left panel corresponds to the Einstein-aether case, while the right one is the khronometric one. The other parameters satisfy the equations \eqref{eq:solar}  coming from Solar System constraints. The areas outside the shaded regions are ruled out by stability/Cherenkov considerations (light blue), BBN (dark orange, to be discussed in Sec.~\ref{sec:backg}) and pulsar constraints (dark purple). The red-dotted line corresponds to the values of parameters for which the orbital decay rate agrees with GR in the weak limit. Observe that the pulsar constraints are much more stringent than the others.
 The figures are taken from \cite{Yagi:2013ava}.
}
\end{center}
\end{figure}

Given the previous results, one can use the  sensitivities of  \cite{Yagi:2013qpa,Yagi:2013ava}
to analyse the change in the orbital period of different binaries within the approximations we
discussed. In \cite{Yagi:2013qpa,Yagi:2013ava} the sensitivities were calculated and presented in terms of the LV parameters and the compactness of the sources, $C^*\equiv G_N \tilde m/R^*$, where $R^*$ is the radius of the source of mass $\tilde m$. These results  were used for three concrete cases:  PSR J1141-6545~\cite{Bhat:2008ck}, PSR J0348+0432~\cite{Antoniadis:2013pzd} and PSR J0737-3039~\cite{Kramer:2006nb}. The first two are composed of a neutron star and white dwarf and the last one is a double binary pulsar (see \cite{Yagi:2013ava} for more details). The very well-studied Hulse-Taylor binary pulsar, PSR 1913+16~\cite{Hulse:1974eb,Taylor:1989sw}, could not be used since it has an eccentricity of roughly $e=0.6$.

Putting altogether, using the constraints from the solitary pulsars and saturating the bounds for $\alpha^{PPN}_1$ and $\alpha^{PPN}_2$ coming from the Solar System tests (cf. Sec.~\ref{sec:solar}) one gets the constraints shown in Fig.~\ref{fig3} \cite{Yagi:2013qpa}. 
 Besides being very stringent, and contrary to Solar System constraints,  these constraints apply to \emph{all} the parameter space. Thus, they bound regions of parameters which are completely untestable in the regimes of weak gravity. In particular, they  constrain the possible deviations of the speed of gravitational waves
with respect to $c$ to be of order ${\cal O}(10^{-1})$. 
  
Before closing this section, let us mention that further tests will be possible once gravitational waves are detected, cf. \cite{Chatziioannou:2012rf}. Of particular importance are the
differences in the wave-form related to  the emission of extra polarizations as compared to GR  \cite{Foster:2006az,Blas:2011zd}
(see \cite{Hansen:2014ewa} for forecasts for testing LV with future observations).
A proper understanding of the signal emitted in LV theory also requires the understanding of rotating solutions for which some results can be found in
\cite{Barausse:2012ny,Wang:2012nv}

\subsection{Cosmology}\label{sec:cosmo}

Finally, let us briefly discuss the consequences for cosmology
 of the presence of an aether field $u^\m$ with action \eqref{ae-action}. For this, we also assume the presence of a dark matter sector
and a dark energy sector, both of which may include couplings to the aether.  As mentioned before, the mechanism yielding the initial conditions for the hot Big Bang (e.g inflation)
can be also modified in LV scenarios.

 Recall that  the existing  laboratory bounds  on LV in the SM are so stringent \cite{Kostelecky:2008ts,Mattingly:2005re}  that they are equivalent, as far as most cosmological consequences are  concerned\footnote{The only place where this approximation may not hold is for extremely 
 energetic phenomena, close or above $M_P$. We thank T. Jacobson for pointing this out.},  to a total decoupling of SM with $u^\m$.  On the other hand, similar restrictions to LV of the dark sector  are non-existent, 
 hence one can  consider arbitrary couplings\footnote{It is worth pointing out that a future direct detection of DM may yield strong constraints
on LV in this sector. Indeed, a direct detection would imply a relatively strong coupling
between the DM and visible sectors. In that case LV would be transferred from DM to
ordinary matter via loop diagrams and thus would be subject to the tight existing bounds
on LV in the SM. Similar arguments were used in  \cite{Bovy:2008gh,Carroll:2008ub}  to constrain violation
of the equivalence principle in the DM sector in the LI context. However, in the absence of
any direct detection so far and since the loop calculations are rather model-dependent, we find
it useful to study the bounds on LV in DM following from its gravitational manifestations,
in particular in cosmology.} of this sector to the field $u^\m$. For example, a collection of DM particles may be coupled to the aether through an action similar to that of Eq.~\eqref{eq:acsens}. A key difference is that the function $\tilde m(\gamma_A)$  now refers to a universal coupling for each particle species (and charge). This generates a long-range interaction in the DM sector which is not present in the SM, hence DM violates the \emph{weak equivalence principle}. The extension of this formalism for more generic situations where DM can be described as a  fluid was presented in 
\cite{Blas:2012vn}. For a perfect fluid without pressure, the action reduces simply to
\be
\label{eq:DM}
S_{[DM]}=-  m_f\int d^4 x \sqrt{-g}\, n\, F(u^\m v_\m), 
\ee
where $m_f$ is the mass of the fluid element,  $n$ is the number density of the fluid and
$v^{\mu}$ is its four-velocity.  In deriving the
equations of motion, the variation of the fields 
must be subject to the constraints\footnote{Alternatively, within the
  so-called pull-back formalism, one introduces a triple of scalar
  fields parameterizing the fluid elements and varies with respect to
  these fields without any restrictions
  \cite{Andersson:2006nr,Blas:2012vn}.} $v^\m v_\m=-1$, 
$\nabla_\m(n\,v^\m)=0$, the latter expressing the particle number
conservation. The resulting equations for the aether-DM system can be
found in \cite{Blas:2012vn}.  Generalizations of the previous construction, in particular with an action for the aether different from \eqref{ae-action}, have also been used to modify GR into a MOND-like theory \cite{Blanchet:2012ub}.
Finally, the projectable Ho\v rava gravity model included an integration constant that could play the role of dark matter \cite{Mukohyama:2010xz}.

Meanwhile LV allows for new alternatives to generate the current accelerated expansion of the universe. 
In particular, for models of dynamical dark energy its coupling to $u^\m$ allows for potentials with remarkable properties. 
An interesting possibility is the $\Theta$CDM model of \cite{Blas:2011en} where one considers the existence of a field $\Theta$, with an action invariant under shift transformations
$\Theta \mapsto \Theta +const.$ and coupled to $u^\m$,
\be\begin{split}
\label{Thetaact}
S_{[\Theta]}=\int d^4x \sqrt{-g}& \bigg(\frac{g^{\m\n}\d_\m\Theta\d_\n\Theta}{2}+
\varkappa\frac{(u^\mu\d_\mu\Theta)^2}{2}
+\mu^2u^\mu\d_\mu\Theta\bigg)\;.
\end{split}
\ee
The LV term proportional to $\m^2$ can drive the current acceleration of the universe and can be easily protected from ultraviolet corrections by the symmetry $\Theta\mapsto -\Theta$. Despite the attractive features of this or similar models, we will not discuss them further in this review and assume a cosmological constant as the source of the current accelerated expansion of the universe (see \cite{Audren:2013dwa} for the current constraints on $\Theta$CDM).  For scenarios with generalised aether action or with LV dynamics different from \eqref{ae-action} see e.g.  \cite{Zuntz:2010jp} and \cite{Libanov:2007mq}.

\subsubsection{Inflation}\label{sec:inflation}

 Cosmological inflation may also be modified in LV scenarios at different levels. 
A first important difference with respect to the $\Lambda$CDM standard model of cosmology (see e.g. 
 \cite{Mukhanov:2005sc} for review) is the interpretation of the homogeneity and isotropy observed in the present Universe. The standard picture of inflation assumes that these two properties come from the exponential expansion of a tiny region of space. This selects a preferred frame in the universe that eventually generates the  CMB frame \cite{Mukhanov:2005sc}. Contrary to the preferred frame of LV this frame does not correspond to a vacuum solution and tends to disappear as the universe expands.  Besides, any choice of frame is identical to any other as a consequence of LI. But in the presence of $u^\m$ with a time-like vev, this field already selects a local frame and  there is no reason \emph{a priori} why the initial ``clock'' of inflation should align with it. This misalignment  would produce an anisotropic universe at large scales. Remarkably, the two frames may align due to the dynamical evolution: it was proven in \cite{Kanno:2006ty,Carruthers:2010ii}
 that  if one introduces anisotropies in the $u^\m$ and metric fields  the isotropic solution is an attractor of the dynamics for accelerating universes. 
 Hence we expect that at the end of inflation, the misalignment of the aether with the  reheating surface will be minuscule  and  it is natural to assume that the background solution is homogeneous and isotropic\footnote{Even more interesting would be to
 allow for slight misalignment and constrain it from observations.}.
The presence of 
  \eqref{ae-action} has a very mild influence in the background evolution in this case (see Sec.~\ref{sec:backg}).

The primordial perturbation spectra are modified   even in the absence of a direct coupling between the inflaton and $u^\m$. In general, the amplitudes of the spectra of all helicities including helicity-2 GW modes are modified although their power law exponent are not in general \cite{Lim:2004js, Wang:2010an, Solomon:2013iza}. In addition, Ref. \cite{ArmendarizPicon:2010rs} showed that there exist unstable growing modes under certain choices of the LV parameters. Ref. \cite{Li:2007vz} undertook a study of the CMB spectra predicted by \cite{Lim:2004js}, without considering any couplings to SM particles and found that the signatures are not highly constraining.
For first results about the bispectrum, see \cite{Chialva:2011iz,Chialva:2011hc}.
The phenomenology of the model is extended once  one considers direct  couplings of $u^\m$ to the inflaton \cite{Kanno:2006ty,Donnelly:2010cr,Solomon:2013iza}.

Other ideas related to LV and inflation  have been discussed in the past. \emph{Ghost inflation} \cite{ArkaniHamed:2003uz} provided an alternative to the standard inflationary paradigm with an spontaneous breaking of Lorentz invariance, which manifests in the modified dispersion relation of the inflaton and very particular shapes of non-gaussianities. This model was extended to high-energies recently in 
\cite{Ivanov:2014yla} by coupling it to the khronometric model. Other modification of the model was the khronometric inflation, where the
inflaton plays the role of khronon during inflation \cite{Creminelli:2012xb}. Finally, it was also noticed that
the modified   dispersion relations \eqref{eq:lvgw}     can generate scale invariant primordial perturbations without inflation \cite{Mukohyama:2010xz}.

In the following, we will focus on late time cosmology for which we will assume the standard Gaussian and adiabatic initial conditions.

\subsubsection{Background evolution}\label{sec:backg}

Once  assuming  isotropy and homogeneity, the   consequences of breaking LI in gravity are very mild. 
To see this, let us consider a model with a cosmological constant, standard matter (not coupled to $u^\m$) and gravity and dark matter actions given by  \eqref{ae-action} and \eqref{eq:DM}. After imposing the standard FLRW form for the metric 
\be
\label{eq:FLRW}
ds^2=a(t)^2 (dt^2-d{\bf x}^2),
\ee
the equations of motion for the scale factor $a(t)$ are
\be
\frac{\dot a^2}{a^4}=\frac{8\pi G_c}{3}\sum_i \rho_i,
\ee
where $\rho_i$ represent the mass densities of the different species of the universe\footnote{Note that
for  DM, even if the action \eqref{eq:DM} modifies the inertial mass, this just corresponds to renormalization of the   the background quantity $\rho_{DM}$ as defined 
by the previous equation.} and
\be
\label{eq:Gc}
G_c=G_N\left(\frac{2-c_a}{2+c_3+c_1+3c_2}\right).
\ee
Thus, at the background level the  aether field  with action \eqref{ae-action} effectively \emph{tracks} the expansion of the universe, and mimics the dynamics of any matter content in the Universe \cite{Carroll:2004ai}. As we will see later, this tracking behavior does not extend to the perturbation level.

The expression \eqref{eq:Gc} implies that the ``Newton's constant'' that weighs the contribution of matter to the cosmic expansion is not the one governing local gravitational physics. As noticed in \cite{Carroll:2004ai}, the production of elements during big bang nucleosynthesis (BBN) is sensitive to $G_c$, and the previous difference can be constrained to 
\be
\left|\frac{G_c}{G_N}-1\right| <0.13.
\ee
It is remarkable that for the Einstein-aether case, the imposition of the Solar System bounds  \eqref{eq:solar}
cancels the leading order difference between $G_c$ and $G_N$ (see Fig.~\ref{fig3}). This means that these models are
harder to test using BBN and this will also be the case for the perturbations discussed in the next section.

\subsubsection{Perturbations}

As discussed in Sec.~\ref{sec:inflation}, the dynamics of perturbations around a homogeneous and isotropic background are modified in LV theories both at the production time and during their evolution. In this section we will focus on late time cosmology and  assume that there was a mechanism that generated an initial power spectrum of Gaussian perturbations in the adiabatic mode. 

Given these initial conditions, the perturbations around the FLRW background \eqref{eq:FLRW} are very sensitive to LV. This comes from different phenomena \cite{Audren:2013dwa,Audren:2014hza}:  first, the theory contains extra light degrees of freedom (a helicity-0 -the khronon- and also a helicity-1 vector for the Einstein-aether case), which are coupled to the metric and (maybe) to DM. Second, $G_c\neq G_N$ which modifies the Poisson equation \cite{Kobayashi:2010eh}. Furthermore, the misalignment of the perturbations with respect to  the aether generates anisotropic stress in the effective energy-momentum tensor. We now briefly discuss these effects, and refer the reader to the discussion (in particular the figures showing the different effects) in \cite{Audren:2013dwa,Audren:2014hza}
for better understanding.

The presence of the extra degrees of freedom  in the energy budget does not have a big impact for the CMB or linear structure formation \cite{Lim:2004js,Zuntz:2008zz,Audren:2013dwa}.  This is because these new species are relativistic (massless and with speeds 
given by  \eqref{eq:speeds}), and they do not cluster. It is still interesting to note that there exist phases with unstable  helicity-1 vector perturbations which may lead to large B-modes in the CMB \cite{ArmendarizPicon:2010rs,Ichiki:2011ah} which may muddy the waters in attempts at detecting primordial GW modes (helicity-2) modes via B-mode polarizations on the CMB (e.g. \cite{Ade:2014xna}).

Meanwhile, the Poisson equation for the gravitational potential $\phi$ 
 is modified even when all matter (including DM and dark energy) is LI,
\be
\label{eq:modPoi}
\Delta \phi=\frac{3}{2}\frac{G_N}{G_c}\left(\frac{\dot a}{a}\right)^2
\sum_i \Omega_i \delta_i,
\ee 
with the standard notation for the density contrasts $\delta_i$ and the $\Omega_i$ density fractions \cite{Audren:2013dwa}. This has several phenomenological consequences: first, it affects the way the acoustic oscillations
in the CMB are produced since it modifies the effective speed of sound waves in the photon-baryon plasma before recombination  \cite{Audren:2013dwa}. Furthermore, during matter domination it modifies the growth rate of cold matter to 
\be
\label{eq:growth}
\delta_{[CM]}\propto a^{\frac{1}{4}\left(-1+\sqrt{1+24 G_N/G_c}\right)}.
\ee
This has an impact both in the linear power spectrum and in the integrated Sachs-Wolf of the CMB.
The previous effects (in particular the shifting of the CMB  peaks) are very characteristic of LV, which means that they 
are not degenerate with parameters of the $\Lambda$CDM model and are easily constrained.

Finally, the presence of anisotropic stress
suppresses the perturbations in the CMB  and  the linear power spectrum at the scales corresponding to the sound speed
of the scalar-mode, $\varv_0$ in \eqref{eq:speeds}.  
 This contribution is
 proportional to $c_\sigma$. This is is expected from general considerations connecting the anisotropic stress and corrections to graviton propagator \cite{Saltas:2014dha}.
 The bounds on anisotropic stress are weaker since its effects are partially degenerate with an overall rescaling of the amplitude of the primordial perturbations\footnote{We thank
 M. Ivanov for discussions about this point.}  \cite{Audren:2013dwa,Audren:2014hza}.  

 It is also interesting to realise that the previous  effects in cosmology (existence of extra species, modification of the growth rate and
anisotropic stress)   are quite generic for theories of  modified gravity \cite{Clifton:2011jh,Amendola:2012ys,Joyce:2014kja}. Thus, all the bounds related to LV theories can be reinterpreted as bounds on generic features of gravity an vice versa.

The addition of couplings of  DM  to the aether, Eq.~\eqref{eq:DM}, has very interesting phenomenological consequences \cite{Blas:2012vn,Audren:2014hza}
To better understand them, one can take the Newtonian limit of \eqref{eq:DM}. We already showed the first step in this direction in  \eqref{taylor-pp}. An important difference is that in  \eqref{taylor-pp} the sensitivities are effective parameters   related to the strong gravitational field of the object while for DM they correspond to fundamental parameters of the theory. The action in the pressure-less perfect fluid approximation  reads
\be
\label{eq:DMY}
S_{[DM]}=\int d^4 x\rho\left[\frac{(V^i)^2}{2}-\phi-Y \frac{(u^i-V^i)^2}{2}\right],
\ee
with $Y\equiv F'(1)$, $V^i$ is the three-velocity of the fluid and $u^i$ the spatial perturbation of the aether. This expression summarises very well the new phenomenology associated to LV in this sector (i.e. $Y\neq 0$). Notice that the source of Newton's potential is not modified. But the kinetic energy (or equivalently the inertial mass) acquires a factor $(1-Y)$  even when $u^i=0$ (the same phenomenon as in \eqref{active-mass}). This modifies the way in which DM  responds to the gravitational field and changes the Jean's instability to 
\be
\label{unscreen}
\delta_{[DM]} \propto
a^{\frac{1}{4}\left(-1+\sqrt{25+\frac{24Y}
{(1-Y)}}\right)}
\,. 
\ee
This anomalous growth is related to the violation of the equivalence principle in the DM sector.
But from \eqref{eq:DMY} we see that the term proportional to $Y$ is minimised if the DM fluid aligns with the aether $V^i=u^i$. 
The way this happens is the following: the action \eqref{eq:DMY} shows that the DM interacts also through a force mediated by $u^i$. 
This new (5th) force is such that for large scales it exactly screens the term proportional to $Y$ in   \eqref{eq:DMY}, while
its effects at short distances are a small correction to the Newtonian potential \cite{Blas:2012vn}. 
The frontier between both regimes happens at the comoving momentum  ($G_0\equiv 
1/(8\pi M_0^2)$)
\be
\label{eq:ky}
k^2_{Y,0}\equiv\frac{3Y\Omega_{dm}H_0^2}{(\b+\l)(1-Y)}\frac{G_0}{\Gc}.
\ee
Notice that this scale is characterised by the LV parameters in \eqref{ae-action}, $Y$
and the matter content of the universe. The presence of $k_Y$ within the regime of observation generates scale dependent features
that allow us to efficiently constrain $Y$. Notice, however, that for very small $\beta+\lambda$, this scale will be very small, 
and the LV effects will move away from the linear regime tested by CMB and the linear power spectrum. This is the way in which the
effects associated to the new  interaction  mediated by $u^i$ are screened.
Finally, it is worth noticing that since DM and baryons now react differently to the gravitational field, a non-vanishing $Y$   yields a new scale dependent bias in the distribution of dark matter as compared with galaxies  \cite{Blas:2012vn}. 

The first comparison of Einstein-aether models  with CMB and galaxy clustering data was presented in \cite{Zuntz:2008zz}. This work assumed $Y=0$ and focused on  the Einstein-aether case. As we mentioned before, once the Solar System constraints are imposed $G_c\approx G_N$ and the anomalous growth in \eqref{eq:growth} is cancelled. Still, we have the effects coming from the anisotropic stress, proportional to $\beta$ ($c_\sigma$). 
Eventually this translates into relatively mild bounds in the model.

The generic comparison with $Y\neq 0$ and including the khronometric case recently appeared in  \cite{Audren:2014hza} (see also \cite{Audren:2013dwa}).
 In that work,  the CMB and the linear power spectrum in universes with LV and adiabatic Gaussian initial conditions were computed
with the Boltzmann code {\sc Class} \cite{Blas:2011rf}. The results were compared with data from the  Planck satellite   \cite{Planck} for the CMB and WiggleZ telescope \cite{WiggleZ} for the linear power spectrum. 
 The regions of allowed parameters were scanned using the Monte Carlo code {\sc Monte Python}  \cite{Audren:2012wb}, which produced 
 the one-dimensional posterior distributions for the different parameters, together with the allowed areas at different levels of confidence.
As for the case of the pulsars, Sec.~\ref{sec:strong}, this analyses  focused in the parameter space allowed by  the Solar System tests   \eqref{eq:solar}.   
The bounds can be found in Table~\ref{table}.  The  results are divided into four different sets:  Einstein-aether vs khronometric and vanishing vs. non-vanishing $Y$.  The first distinction 
is important because in the Einstein-aether case, $G_N\approx G_c$ and the bounds are significantly milder. The second distinction is 
relevant because if $k_Y$ in \eqref{eq:ky} is higher than the scales relevant for the observations, $Y$ remains basically unconstrained, and
the bayesian analyses in this case is very different from the case $Y=0$ \cite{Audren:2014hza}.
 Notice that the results are given in terms of $c_\chi^2\equiv (\beta+\lambda)/\alpha$, which corresponds to the velocity of the scalar mode $\varv_0$ at small values of the parameters. 
\begin{table*}[t]
\begin{center}
  \begin{tabular}{|lccc|}
     \hline
     & $\alpha$  &
     $c_{\chi}^2$  & $Y$ \\ \hline 
     Einstein-aether \; &
     $<5.0 \cdot 10^{-3}$ & $<240$ & $<0.028$$$ \\[0.2cm]
          khronometric \; & 
     $<1.1 \cdot 10^{-3}$ & $<55$ & $<0.029$$$ \\[0.2cm]
      Einstein-aether, $Y\equiv 0$ \; & 
     $<1.0 \cdot 10^{-2}$ & $<427$ & $-$$$ \\[0.2cm]
       khronometric, $Y\equiv 0$  \; & 
     $<1.8 \cdot 10^{-3}$ & $<91$ & $-$$$ \\[0.2cm]
     \hline
   \end{tabular}
\caption{Upper limits of the LV parameters at 95\% CL  derived from 
the study of the CMB and the linear power spectrum in \cite{Audren:2014hza}. 
These bounds are found after  imposing the Solar System constraints.
}
\label{table}
\end{center}
\end{table*}

Remarkably, cosmological data strongly constrains \emph{all} the parameter space at the same level, and in certain cases even better, than the pulsar constraints. Furthermore, the analysis of \cite{Audren:2014hza} constitutes the first bounds in the possibility of DM violating LI.  

\section{Summary and discussion}

In this short review we have described different phenomenological aspects of violation of Lorentz invariance in the gravitational sector. The motivation to study them comes from quantum gravity, but also as a check of the assumptions of GR. We have focused in the case where Lorentz invariance is broken by the existence of a preferred time direction in every point of space-time. To describe this situation we introduced a 4-vector field $u^\m$ with a time-like vev. The generic theory for this situation is known as Einstein-aether theory. One can also reduce the study to  the case when this field is orthogonal to a family of hyper-surfaces \eqref{eq:kh} (khronometric case). Ho\v rava gravity can be understood as a UV completion of the latter.

An important feature of LV modifications is that their consequences may appear at basically all scales. For short scales (characterised by  energy scales above the value $M_{LV}$ defined in \eqref{MLV}) one expects them to completely change GR. 
We summarised their possible effects in this regime by their impact on the dispersion relations of propagating degrees of freedom \eqref{eq:lvgw} and deviations from Newton's law in the potentials \eqref{eq:new}. These are expected to be the leading LV modifications in the linear regime.  Both cases are characterised by an energy scale related to $M_{LV}$ and we discussed how to bound this scale by current experiments, together with ways to get further constraints. It is interesting to notice that in the Ho\v rava gravity case the potentials are better behaved at short distances with respect to GR. This suggests that singularities may be resolved in this theory of gravity but  more work is needed to clarify this point.

For energy scales smaller than $M_{LV}$ (large distances), the presence of a preferred frame -- both the fixed background and its excitations -- has an impact in all the classical tests of gravity. To study them, we first assumed that the fields in the SM do not couple to $u^\m$ (as shown by laboratory tests). Thus, only the gravitational sector (including the dark matter and the dark energy) are allowed to directly interact with the preferred frame. Very strong bounds on this interaction 
can be found from the study of gravity in the Solar System. Indeed, the presence of $u^\m$ modifies the post-Newtonian dynamics of gravitating systems, and we described in Sec.~\ref{sec:solar} the phenomenological consequences in terms of PPN parameters. 

The fact that gravitons  interact  with a new field with massless excitations ($u^\m$) implies the violation of the strong equivalence principle (gravitons do not follow geodesics of any metric). This has very important consequences in the study of compact objects, where the gravitational energy is non-negligible. In particular, this 
affects the metric generated by neutron stars, and one can use data from different solitary pulsars to put constraints on the possible interaction of gravity with the LV field, cf. Fig.~\ref{fig2}.  Furthermore,  the violation of the strong equivalence principle completely changes the pattern of emission of gravitational waves in binary systems. 
In particular, there is a dipolar component in the radiation which is enhanced with respect to the GR quadrupole by a factor ${\mathcal O}(v^2/c^2)$, where $v$ is the mutual velocity of the binary. This is efficient for systems which are compact enough and with an important dipole (e.g. asymmetric). For these cases, the study of the  emission of dipolar radiation in binary systems puts very stringent constraints on LV in gravity, Fig.~\ref{fig3}. 

Finally, all the cosmic history of the universe may be affected by the presence of a preferred frame. The first consequences happen already at the inflationary stage, where this may generate some degree of anisotropy. At any event, one can assume that inflation happened in a way similar to the standard one in $\Lambda$CDM and consider Gaussian and adiabatic initial conditions. The later cosmological evolution is also vey much affected by LV. 
Even in the case where only the metric is coupled to $u^\m$ we discovered that the
properties of the CMB (positions and relative heights of the peaks) and of the linear power spectrum of matter (related to the growth rate) are modified and can be used to test LV. Furthermore, one can also consider the situation where dark energy or DM couple to $u^\m$. The former case is interesting because LV allows to consider new natural potentials that could be behind the current accelerated expansion of the universe. As far as DM is concerned, the LV effects can be summarised as a long-range 5th force with a screening mechanism. This has a very peculiar effect on CMB and power spectrum, related to the violation of the weak equivalence principle. The bounds from linear cosmology are presented in Table~\ref{table}.

Many of the effects we described present open alleys that deserve further study\footnote{Even if not directly related to gravitation, it is also important to better understand the mechanisms that may generate a SM with very high degree of LI, while allowing for LV in gravity. We briefly mentioned some possibilities at the end of Sec.~\ref{sec:introd}.} . For the behavior of the theory at short distances 
a crucial question is whether the LV proposals (e.g. Ho\v rava gravity) can be complete theories. To prove this, one would like to go beyond scaling arguments and prove the renormalizability of the theory and the structure (or absence) of singularities. As should be clear from the discussion in Sec.~\ref{sec:short}, the study of GW with modified dispersion relations also deserves further clarifications. In particular, one would like to understand the connection to observables that may be available in the near future. On a more mathematical level, it would be interesting to clarify if the initial value formulation of the theory is well-posed. 

Concerning the phenomenology at large distances, the most exciting open questions come from the study of compact objects, gravitational waves and cosmology. 
As we have seen, the gravitational field of a single pulsar-white
dwarf binary, PSR J1738+0333 cf. Fig.~\ref{fig2}, puts very strong constraints on possible deviations from GR. The calculations that we summarised in this review in this direction do not describe all the possible variety of compact objects that have been detected. 
Similarly, the gravitational collapse is not completely understood in LV theories of gravity (see \cite{Garfinkle:2007bk,Barausse:2013nwa} for early work). 
For gravitational waves, a complete characterisation of the wave-form from different sources and further study  in the lines of 
 \cite{Sampson:2014qqa} would be extremely  interesting. This may  require to consider slowly rotating stars, which by itself represents a way to test  theories beyond GR \cite{Pani:2014jra}.
    
Finally, the consequences of LV in cosmology still require further study both at early and late times. For the primordial universe, the  situation where the energies are so high to excite generate gravitational modes close to or above $M_{LV}$  have not been explored in detail. Furthermore, as mentioned in the text, one would like to have a better picture of the way in which the   expected anisotropy  from the misalignment of the inflationary frame with $u^\m$ disappears with time. The naive expectation is that some level of anisotropy could remain in the largest scales, but more clarification is needed  (see e.g. \cite{Kim:2013gka,Carroll:2008br} for related work in anisotropic scenarios). 
At later cosmological times, the most important question is to clarify the influence of LV in non-linear cosmology (e.g. when the clustering of dark matter generates big density perturbations). Of particular important is the fact that the theories under study possess extra light degrees of freedom that can be understood as 5th forces. 
For these forces to directly affect the DM clustering we  require $Y\neq0$. If this charge is not universal, this implies the violation of the equivalence principle
whose influence for non-linear cosmology may be very important  (see e.g. \cite{Blas:2014hya,Gubser:2004du,Kesden:2006vz,Rocha:2012jg,Vogelsberger:2012ku} for different flavours of this possibility). In this respect, the ideal would be to develop both N-body simulations of the theories with LV together with an analytical understanding of the main effects.  Finally, the possibility of DM self-interaction has been extensively studied for smaller (cluster) scales, see e.g. \cite{Kaplinghat:2013kqa}.
To our knowledge the interactions mediated by $u^\m$ (corresponding to large distance interactions) have not been considered, and it would be 
interesting to understand whether the  bounds on $Y$ (self-interaction) coming from halo
shapes and Bullet Cluster bounds are competitive with those presented above. 

 To end this short review, we hope to have convinced the reader that LV theories of gravity are both well motivated theoretically and have phenomenological implications that teach us about the unique features of GR and consistent ways to modify it. This is very important in the age where the data coming from GW and cosmological observations will allow us to push our understanding of gravitation to a new level of precision.

\section*{Acknowledgements}

We are grateful to M. Ivanov, T. Jacobson, S. Liberati and S. Sibiryakov for their very useful discussions and their comments on the draft. 
We are also grateful to E. Barausse, K. Yagi and N. Yunes for clarifications about the effects of LV in the strong gravity regime. EAL is supported by STFC AGP grant ST/L000717/1.



\begin{thebibliography}{99}


\bibitem{Ackerman:2007nb} 
L.~Ackerman, S.~M.~Carroll and M.~B.~Wise,
Phys.\ Rev.\ D {\bf 75}, 083502 (2007)
[Erratum-ibid.\ D {\bf 80}, 069901 (2009)]
[astro-ph/0701357].

\bibitem{Cohen:2006ky} 
A.~G.~Cohen and S.~L.~Glashow,
Phys.\ Rev.\ Lett.\  {\bf 97}, 021601 (2006)
[hep-ph/0601236].

\bibitem{Bailey:2006fd}
  Q.~G.~Bailey and V.~A.~Kostelecky,
  Phys.\ Rev.\ D {\bf 74} (2006) 045001
  [gr-qc/0603030].

\bibitem{Tasson:2014dfa}
  J.~D.~Tasson,
  Rept.\ Prog.\ Phys.\  {\bf 77} (2014) 062901
  [arXiv:1403.7785 [hep-ph]].
  
\bibitem{Cheng:2006us}
  H.~C.~Cheng, M.~A.~Luty, S.~Mukohyama and J.~Thaler,
  JHEP {\bf 0605} (2006) 076
  [hep-th/0603010].

\bibitem{Horava:2009uw}
P.~Horava,
Phys.\ Rev.\ D {\bf 79} (2009) 084008
[arXiv:0901.3775 [hep-th]].

\bibitem{Jacobson:2000xp}
  T.~Jacobson and D.~Mattingly,
  Phys.\ Rev.\ D {\bf 64} (2001) 024028
  [gr-qc/0007031].

\bibitem{Gasperini:1987nq}
  M.~Gasperini,
  Class.\ Quant.\ Grav.\  {\bf 4} (1987) 485.
 
\bibitem{Gasperini:1998eb}
  M.~Gasperini,
  Gen.\ Rel.\ Grav.\  {\bf 30} (1998) 1703
  [gr-qc/9805060].



\bibitem{Blas:2010hb}
D.~Blas, O.~Pujolas and S.~Sibiryakov,
JHEP {\bf 1104} (2011) 018
[arXiv:1007.3503 [hep-th]].


\bibitem{Blas:2009yd}
D.~Blas, O.~Pujolas and S.~Sibiryakov,
JHEP {\bf 0910} (2009) 029
[arXiv:0906.3046 [hep-th]].

\bibitem{Blas:2011ni}
  D.~Blas and S.~Sibiryakov,
  Phys.\ Rev.\ D {\bf 84} (2011) 124043
  [arXiv:1110.2195 [hep-th]].


\bibitem{Tolman:Book}
R.~C.~Tolman,
  ``The Theory of the Relativity of Motion,''
{\it  Univ. of California Press, (1917). 240 p.}

\bibitem{Jacobson:2004ts}
  T.~Jacobson and D.~Mattingly,
  Phys.\ Rev.\  D {\bf 70} (2004) 024003
  [arXiv:gr-qc/0402005].


\bibitem{Mattingly:2005re}
  D.~Mattingly,
  Living Rev.\ Rel.\  {\bf 8} (2005) 5
  [gr-qc/0502097].

\bibitem{Babichev:2007dw}
  E.~Babichev, V.~Mukhanov and A.~Vikman,
  JHEP {\bf 0802} (2008) 101
  [arXiv:0708.0561 [hep-th]].

\bibitem{Donnelly:2010qd}
  W.~Donnelly and T.~Jacobson,
  Phys.\ Rev.\ D {\bf 82} (2010) 081501
  [arXiv:1008.4351 [hep-th]].

\bibitem{Blas:2014ira}
  D.~Blas and S.~Sibiryakov,
  Zh.\ Eksp.\ Teor.\ Fiz.\  {\bf 147} (2015) 578
   [J.\ Exp.\ Theor.\ Phys.\  {\bf 120} (2015) 3,  509]
  [arXiv:1410.2408 [hep-th]].

\bibitem{Blas:2009qj}
D.~Blas, O.~Pujolas and S.~Sibiryakov,
Phys.\ Rev.\ Lett.\ {\bf 104} (2010) 181302
[arXiv:0909.3525 [hep-th]].
  

\bibitem{Jacobson:2013xta}
T.~Jacobson,
``Undoing the Twist: the Ho\v rava Limit of Einstein-Aether,''
arXiv:1310.5115 [gr-qc].


\bibitem{Blas:2009ck}
  D.~Blas, O.~Pujolas and S.~Sibiryakov,
  Phys.\ Lett.\ B {\bf 688} (2010) 350
  [arXiv:0912.0550 [hep-th]].

\bibitem{Papazoglou:2009fj}
  A.~Papazoglou and T.~P.~Sotiriou,
  Phys.\ Lett.\ B {\bf 685} (2010) 197
  [arXiv:0911.1299 [hep-th]].
  

\bibitem{Kostelecky:2008ts}
V.~A.~Kostelecky and N.~Russell,
Rev.\ Mod.\ Phys.\ {\bf 83} (2011) 11
[arXiv:0801.0287 [hep-ph]].


\bibitem{Liberati:2013xla}
  S.~Liberati,
  Class.\ Quant.\ Grav.\  {\bf 30} (2013) 133001
  [arXiv:1304.5795 [gr-qc]].

\bibitem{GrootNibbelink:2004za}
  S.~Groot Nibbelink and M.~Pospelov,
  Phys.\ Rev.\ Lett.\  {\bf 94} (2005) 081601
  [hep-ph/0404271].

  
\bibitem{Pujolas:2011sk}
  O.~Pujolas and S.~Sibiryakov,
  JHEP {\bf 1201} (2012) 062
  [arXiv:1109.4495 [hep-th]].
  
  
\bibitem{Pospelov:2010mp}
  M.~Pospelov and Y.~Shang,
  Phys.\ Rev.\ D {\bf 85} (2012) 105001
  [arXiv:1010.5249 [hep-th]].
  
  
\bibitem{Bednik:2013nxa}
  G.~Bednik, O.~Pujolˆs and S.~Sibiryakov,
  JHEP {\bf 1311} (2013) 064
  [arXiv:1305.0011 [hep-th]].

\bibitem{Chkareuli:2001xe}
  J.~L.~Chkareuli, C.~D.~Froggatt and H.~B.~Nielsen,
  Phys.\ Rev.\ Lett.\  {\bf 87} (2001) 091601
  [hep-ph/0106036].

\bibitem{Mirshekari:2011yq}
S.~Mirshekari, N.~Yunes and C.~M.~Will,
Phys.\ Rev.\ D {\bf 85} (2012) 024041
[arXiv:1110.2720 [gr-qc]].


\bibitem{Yagi:2013ava}
  K.~Yagi, D.~Blas, E.~Barausse and N.~Yunes,
  Phys.\ Rev.\ D {\bf 89} (2014) 084067
  [arXiv:1311.7144 [gr-qc]].

  
\bibitem{Lim:2004js}
  E.~A.~Lim,
  Phys.\ Rev.\ D {\bf 71} (2005) 063504
  [astro-ph/0407437].

\bibitem{Mukohyama:2010xz}
  S.~Mukohyama,
  Class.\ Quant.\ Grav.\  {\bf 27} (2010) 223101
  [arXiv:1007.5199 [hep-th]].
  
  
\bibitem{Wang:2010an}
  A.~Wang,
  Phys.\ Rev.\ D {\bf 82} (2010) 124063
  [arXiv:1008.3637 [hep-th]].
  
\bibitem{Brandenberger:2012aj}
  R.~H.~Brandenberger and J.~Martin,
  Class.\ Quant.\ Grav.\  {\bf 30} (2013) 113001
  [arXiv:1211.6753 [astro-ph.CO]].

\bibitem{Barrow:2014fsa}
  J.~D.~Barrow, M.~Lagos and J.~Magueijo,
  Phys.\ Rev.\ D {\bf 89} (2014) 083525
  [arXiv:1401.7491 [astro-ph.CO]].

  \bibitem{Adshead:2009bz} 
P.~Adshead and E.~A.~Lim,
Phys.\ Rev.\ D {\bf 82}, 024023 (2010)
[arXiv:0912.1615 [astro-ph.CO]].
  

\bibitem{Maldacena:2002vr}
J.~M.~Maldacena,
JHEP {\bf 0305} (2003) 013
[astro-ph/0210603].
   

\bibitem{Bailey:2014bta}
  Q.~G.~Bailey, A.~Kostelecky and R.~Xu,
  arXiv:1410.6162 [gr-qc].

\bibitem{Kapner:2006si}
  D.~J.~Kapner, T.~S.~Cook, E.~G.~Adelberger, J.~H.~Gundlach, B.~R.~Heckel, C.~D.~Hoyle and H.~E.~Swanson,
  Phys.\ Rev.\ Lett.\  {\bf 98} (2007) 021101
  [hep-ph/0611184].


  
\bibitem{Murata:2014nra}
  J.~Murata and S.~Tanaka,
  ``Review of short-range gravity experiments in the LHC era,''
  arXiv:1408.3588 [hep-ex].

\bibitem{Will:2014xja}
  C.~M.~Will,
  Living Rev.\ Rel.\  {\bf 17} (2014) 4
  [arXiv:1403.7377 [gr-qc]].

\bibitem{Calcagni:2009ar}
  G.~Calcagni,
  JHEP {\bf 0909} (2009) 112
  [arXiv:0904.0829 [hep-th]].

\bibitem{Kiritsis:2009sh}
  E.~Kiritsis and G.~Kofinas,
  Nucl.\ Phys.\ B {\bf 821} (2009) 467
  [arXiv:0904.1334 [hep-th]].

\bibitem{Brandenberger:2009yt}
  R.~Brandenberger,
  Phys.\ Rev.\ D {\bf 80} (2009) 043516
  [arXiv:0904.2835 [hep-th]].


\bibitem{Brauner:2010wm}
  T.~Brauner,
  Symmetry {\bf 2} (2010) 609
  [arXiv:1001.5212 [hep-th]].
  
\bibitem{Nicolis:2012vf}
  A.~Nicolis and F.~Piazza,
  Phys.\ Rev.\ Lett.\  {\bf 110} (2013) 011602
  [arXiv:1204.1570 [hep-th]].
  
\bibitem{Watanabe:2012hr}
  H.~Watanabe and H.~Murayama,
  Phys.\ Rev.\ Lett.\  {\bf 108} (2012) 251602
  [arXiv:1203.0609 [hep-th]].

\bibitem{Elliott:2005va} 
  J.~W.~Elliott, G.~D.~Moore and H.~Stoica,
  JHEP {\bf 0508}, 066 (2005)
  [hep-ph/0505211].
  
\bibitem{Eling:2005zq}
  C.~Eling,
  Phys.\ Rev.\ D {\bf 73} (2006) 084026
   [Erratum-ibid.\ D {\bf 80} (2009) 129905]
  [gr-qc/0507059].

\bibitem{Garfinkle:2011iw}
D.~Garfinkle and T.~Jacobson,
Phys.\ Rev.\ Lett.\ {\bf 107} (2011) 191102
[arXiv:1108.1835 [gr-qc]].


\bibitem{Blanchet:2006zz}
L.~Blanchet,
Living Rev.\ Rel.\  {\bf 9} (2006) 4.

\bibitem{Will:Book}
  C.~M.~Will,
  ``Theory and experiment in gravitational physics,''
{\it  Cambridge, UK: Univ. Pr. (1993), 380 p.}

\bibitem{Wein:Book}
  S.~Weinberg, 
   ``Gravitation and Cosmology: Principles and Applications of the General Theory of Relativity,''
{\it John Wiley \& Sons, Inc. (1972), 657 p. }



\bibitem{Kanno:2006ty}
  S.~Kanno and J.~Soda,
  Phys.\ Rev.\ D {\bf 74} (2006) 063505
  [hep-th/0604192].

\bibitem{Carruthers:2010ii}
  I.~Carruthers and T.~Jacobson,
  Phys.\ Rev.\ D {\bf 83} (2011) 024034
  [arXiv:1011.6466 [gr-qc]].
  

\bibitem{Foster:2005dk}
B.~Z.~Foster and T.~Jacobson,
Phys.\ Rev.\  D {\bf 73} (2006) 064015
[arXiv:gr-qc/0509083].


\bibitem{Blas:2011zd}
D.~Blas and H.~Sanctuary,
Phys.\ Rev.\ D {\bf 84} (2011) 064004
[arXiv:1105.5149 [gr-qc]].

 
\bibitem{Will:2014bqa}
  C.~M.~Will,
  ``Was Einstein Right? A Centenary Assessment,''
  arXiv:1409.7871 [gr-qc].

  
 
\bibitem{Foster:2007gr}
  B.~Z.~Foster,
  Phys.\ Rev.\  D {\bf 76} (2007) 084033
  [arXiv:0706.0704 [gr-qc]].

\bibitem{Damour:1992we}
  T.~Damour and G.~Esposito-Farese,
  Class.\ Quant.\ Grav.\  {\bf 9} (1992) 2093.
  
\bibitem{Damour:1992ah}
  T.~Damour and G.~Esposito-Farese,
  Phys.\ Rev.\ D {\bf 46} (1992) 4128.
 
\bibitem{Weisberg:2013pha}
  J.~M.~Weisberg,
  ``Using Binary Pulsars to Test Lorentz Symmetry in the Gravitational Sector,''
  arXiv:1310.7309 [gr-qc].
  
\bibitem{Shao:2013wga}
  L.~Shao, R.~N.~Caballero, M.~Kramer, N.~Wex, D.~J.~Champion and A.~Jessner,
  Class.\ Quant.\ Grav.\  {\bf 30} (2013) 165019
  [arXiv:1307.2552 [gr-qc]].
  
\bibitem{Shao:2012eg}
  L.~Shao and N.~Wex,
  Class.\ Quant.\ Grav.\  {\bf 29} (2012) 215018
  [arXiv:1209.4503 [gr-qc]].
 
\bibitem{Freire:2012mg}
  P.~C.~C.~Freire, N.~Wex, G.~Esposito-Farese, J.~P.~W.~Verbiest, M.~Bailes, B.~A.~Jacoby, M.~Kramer and I.~H.~Stairs {\it et al.},
  Mon.\ Not.\ Roy.\ Astron.\ Soc.\  {\bf 423} (2012) 3328
  [arXiv:1205.1450 [astro-ph.GA]].
  
\bibitem{Yagi:2013qpa}
  K.~Yagi, D.~Blas, N.~Yunes and E.~Barausse,
  Phys.\ Rev.\ Lett.\  {\bf 112} (2014) 161101
  [arXiv:1307.6219 [gr-qc]].
  
\bibitem{Eling:2007xh}
  C.~Eling, T.~Jacobson and M.~Coleman Miller,
  Phys.\ Rev.\ D {\bf 76} (2007) 042003
   [Erratum-ibid.\ D {\bf 80} (2009) 129906]
  [arXiv:0705.1565 [gr-qc]].


  


\bibitem{Barausse:2011pu}
  E.~Barausse, T.~Jacobson and T.~P.~Sotiriou,
  Phys.\ Rev.\ D {\bf 83} (2011) 124043
  [arXiv:1104.2889 [gr-qc]].
  
\bibitem{Garfinkle:2007bk} 
D.~Garfinkle, C.~Eling and T.~Jacobson,
Phys.\ Rev.\ D {\bf 76}, 024003 (2007)
[gr-qc/0703093].

\bibitem{Barausse:2013nwa}
  E.~Barausse and T.~P.~Sotiriou,
  Class.\ Quant.\ Grav.\  {\bf 30} (2013) 244010
  [arXiv:1307.3359 [gr-qc]].
  
  
\bibitem{Hulse:1974eb}
  R.~A.~Hulse and J.~H.~Taylor,
  Astrophys.\ J.\  {\bf 195} (1975) L51.
 
\bibitem{Taylor:1989sw}
  J.~H.~Taylor and J.~M.~Weisberg,
  Astrophys.\ J.\  {\bf 345} (1989) 434.
  
  
\bibitem{Foster:2006az}
  B.~Z.~Foster,
  Phys.\ Rev.\  {\bf D73}, 104012 (2006).
  [gr-qc/0602004]. 
  

 
\bibitem{Kramer:2006nb}
  M.~Kramer, I.~H.~Stairs, R.~N.~Manchester, M.~A.~McLaughlin, A.~G.~Lyne, R.~D.~Ferdman, M.~Burgay and D.~R.~Lorimer {\it et al.},
  Science {\bf 314} (2006) 97
  [astro-ph/0609417].
 
  
\bibitem{Bhat:2008ck}
  N.~D.~R.~Bhat, M.~Bailes and J.~P.~W.~Verbiest,
  Phys.\ Rev.\ D {\bf 77} (2008) 124017
  [arXiv:0804.0956 [astro-ph]].
  
  
 
\bibitem{Antoniadis:2013pzd}
  J.~Antoniadis, P.~C.~C.~Freire, N.~Wex, T.~M.~Tauris, R.~S.~Lynch, M.~H.~van Kerkwijk, M.~Kramer and C.~Bassa {\it et al.},
  Science {\bf 340} (2013) 6131
  [arXiv:1304.6875 [astro-ph.HE]].
  
  

\bibitem{Chatziioannou:2012rf}
  K.~Chatziioannou, N.~Yunes and N.~Cornish,
  Phys.\ Rev.\ D {\bf 86} (2012) 022004
  [arXiv:1204.2585 [gr-qc]].
  
\bibitem{Hansen:2014ewa}
  D.~Hansen, N.~Yunes and K.~Yagi,
  Phys.\ Rev.\ D {\bf 91} (2015) 8,  082003
  [arXiv:1412.4132 [gr-qc]].
  
\bibitem{Barausse:2012ny}
  E.~Barausse and T.~P.~Sotiriou,
  Phys.\ Rev.\ Lett.\  {\bf 109} (2012) 181101
   [Erratum-ibid.\  {\bf 110} (2013) 039902]
  [arXiv:1207.6370].
 
\bibitem{Wang:2012nv}
  A.~Wang,
  Phys.\ Rev.\ Lett.\  {\bf 110} (2013) 091101
  [arXiv:1212.1876].
  
\bibitem{Bovy:2008gh}
  J.~Bovy and G.~R.~Farrar,
  Phys.\ Rev.\ Lett.\  {\bf 102} (2009) 101301
  [arXiv:0807.3060 [hep-ph]].
  
\bibitem{Carroll:2008ub}
  S.~M.~Carroll, S.~Mantry, M.~J.~Ramsey-Musolf and C.~W.~Stubbs,
  Phys.\ Rev.\ Lett.\  {\bf 103} (2009) 011301
  [arXiv:0807.4363 [hep-ph]].

\bibitem{Blas:2012vn}
  D.~Blas, M.~M.~Ivanov and S.~Sibiryakov,
  JCAP {\bf 1210} (2012) 057
  [arXiv:1209.0464 [astro-ph.CO]].
   
  
\bibitem{Andersson:2006nr}
  N.~Andersson and G.~L.~Comer,
  Living Rev.\ Rel.\  {\bf 10} (2007) 1
  [gr-qc/0605010].

       
\bibitem{Blanchet:2012ub}
  L.~Blanchet and S.~Marsat,
  ``Relativistic MOND theory based on the Khronon scalar field,''
  arXiv:1205.0400 [gr-qc].
  

  
\bibitem{Blas:2011en}
  D.~Blas and S.~Sibiryakov,
  JCAP {\bf 1107} (2011) 026
  [arXiv:1104.3579 [hep-th]].
  
  
 
\bibitem{Audren:2013dwa}
  B.~Audren, D.~Blas, J.~Lesgourgues and S.~Sibiryakov,
  JCAP {\bf 1308} (2013) 039
  [arXiv:1305.0009 [astro-ph.CO], arXiv:1305.0009].
 

\bibitem{Zuntz:2010jp}
  J.~Zuntz, T.~G.~Zlosnik, F.~Bourliot, P.~G.~Ferreira and G.~D.~Starkman,
  Phys.\ Rev.\ D {\bf 81} (2010) 104015
  [arXiv:1002.0849 [astro-ph.CO]].
  
\bibitem{Libanov:2007mq}
  M.~Libanov, V.~Rubakov, E.~Papantonopoulos, M.~Sami and S.~Tsujikawa,
  JCAP {\bf 0708} (2007) 010
  [arXiv:0704.1848 [hep-th]].
  

\bibitem{Mukhanov:2005sc}
  V.~Mukhanov,
  ``Physical foundations of cosmology,''
  {\it Cambridge, UK: Univ. Pr. (2005) 421 p.}

\bibitem{Solomon:2013iza}
  A.~R.~Solomon and J.~D.~Barrow,
  Phys.\ Rev.\ D {\bf 89} (2014) 024001
  [arXiv:1309.4778 [astro-ph.CO]].
 

\bibitem{ArmendarizPicon:2010rs}
  C.~Armendariz-Picon, N.~F.~Sierra and J.~Garriga,
  JCAP {\bf 1007} (2010) 010
  [arXiv:1003.1283 [astro-ph.CO]].

\bibitem{Li:2007vz}
  B.~Li, D.~Fonseca Mota and J.~D.~Barrow,
  Phys.\ Rev.\ D {\bf 77} (2008) 024032
  [arXiv:0709.4581 [astro-ph]].
  
  

\bibitem{Chialva:2011iz}
  D.~Chialva,
  JCAP {\bf 1201} (2012) 037
  [arXiv:1106.0040 [hep-th]].

\bibitem{Chialva:2011hc}
  D.~Chialva,
  JCAP {\bf 1210} (2012) 037
  [arXiv:1108.4203 [astro-ph.CO]].
   
 
\bibitem{Donnelly:2010cr} 
W.~Donnelly and T.~Jacobson,
Phys.\ Rev.\ D {\bf 82}, 064032 (2010)
[arXiv:1007.2594 [gr-qc]].
 
 
\bibitem{ArkaniHamed:2003uz}
  N.~Arkani-Hamed, P.~Creminelli, S.~Mukohyama and M.~Zaldarriaga,
  JCAP {\bf 0404} (2004) 001
  [hep-th/0312100].
  
\bibitem{Ivanov:2014yla}
  M.~M.~Ivanov and S.~Sibiryakov,
  JCAP {\bf 1405} (2014) 045
  [arXiv:1402.4964 [astro-ph.CO]].
  
\bibitem{Creminelli:2012xb}
  P.~Creminelli, J.~Norena, M.~Pena and M.~Simonovic,
  JCAP {\bf 1211} (2012) 032
  [arXiv:1206.1083 [hep-th]].
  
   
\bibitem{Carroll:2004ai}
  S.~M.~Carroll and E.~A.~Lim,
  Phys.\ Rev.\ D {\bf 70} (2004) 123525
  [hep-th/0407149].
  
    
\bibitem{Audren:2014hza}
  B.~Audren, D.~Blas, M.~M.~Ivanov, J.~Lesgourgues and S.~Sibiryakov,
  ``Cosmological constraints on deviations from Lorentz invariance in gravity and dark matter,''
  arXiv:1410.6514 [astro-ph.CO].
  
  
 
\bibitem{Kobayashi:2010eh}
  T.~Kobayashi, Y.~Urakawa and M.~Yamaguchi,
  JCAP {\bf 1004} (2010) 025
  [arXiv:1002.3101 [hep-th]].
 
  
  
\bibitem{Zuntz:2008zz} 
  J.~A.~Zuntz, P.~G.~Ferreira and T.~G.~Zlosnik,
  Phys.\ Rev.\ Lett.\  {\bf 101}, 261102 (2008)
  [arXiv:0808.1824 [gr-qc]].
 
 
\bibitem{Ichiki:2011ah} 
  K.~Ichiki, K.~Takahashi and N.~Sugiyama,
      Phys.\ Rev.\ D {\bf 85}, 043009 (2012)
        [arXiv:1112.4705 [astro-ph.CO]].


\bibitem{Ade:2014xna} 
P.~A.~R.~Ade {\it et al.}  [BICEP2 Collaboration],
Phys.\ Rev.\ Lett.\  {\bf 112}, 241101 (2014)
[arXiv:1403.3985 [astro-ph.CO]].
              
\bibitem{Saltas:2014dha}
  I.~D.~Saltas, I.~Sawicki, L.~Amendola and M.~Kunz,
  Phys.\ Rev.\ Lett.\  {\bf 113} (2014) 19,  191101
  [arXiv:1406.7139 [astro-ph.CO]].
    
\bibitem{Clifton:2011jh} 
  T.~Clifton, P.~G.~Ferreira, A.~Padilla and C.~Skordis,
  Phys.\ Rept.\  {\bf 513}, 1 (2012)
  [arXiv:1106.2476 [astro-ph.CO]].

\bibitem{Amendola:2012ys} 
  L.~Amendola {\it et al.}  [Euclid Theory Working Group Collaboration],
  Living Rev.\ Rel.\  {\bf 16}, 6 (2013)
  [arXiv:1206.1225 [astro-ph.CO]].

\bibitem{Joyce:2014kja}
  A.~Joyce, B.~Jain, J.~Khoury and M.~Trodden,
  arXiv:1407.0059 [astro-ph.CO].

\bibitem{Blas:2011rf}
D.~Blas, J.~Lesgourgues and T.~Tram,
JCAP {\bf 1107} (2011) 034
[arXiv:1104.2933 [astro-ph.CO]].


  
  
\bibitem{Planck}
  P.~A.~R.~Ade {\it et al.}  [Planck Collaboration],
  arXiv:1303.5062 [astro-ph.CO].
  
\bibitem{WiggleZ}
  D.~Parkinson, S.~Riemer-Sorensen, C.~Blake, G.~B.~Poole, T.~M.~Davis, S.~Brough, M.~Colless and C.~Contreras {\it et al.},
  Phys.\ Rev.\ D {\bf 86}, 103518 (2012)
  [arXiv:1210.2130 [astro-ph.CO]].
  

\bibitem{Audren:2012wb} 
  B.~Audren, J.~Lesgourgues, K.~Benabed and S.~Prunet,
  JCAP {\bf 1302}, 001 (2013)
  [arXiv:1210.7183 [astro-ph.CO]].
  
\bibitem{Sampson:2014qqa}
  L.~Sampson, N.~Yunes, N.~Cornish, M.~Ponce, E.~Barausse, A.~Klein, C.~Palenzuela and L.~Lehner,
  ``Projected Constraints on Scalarization with Gravitational Waves from Neutron Star Binaries,''
  arXiv:1407.7038 [gr-qc].
  

\bibitem{Pani:2014jra}
  P.~Pani and E.~Berti,
  Phys.\ Rev.\ D {\bf 90} (2014) 024025
  [arXiv:1405.4547 [gr-qc]].


\bibitem{Kim:2013gka}
  J.~Kim and E.~Komatsu,
  Phys.\ Rev.\ D {\bf 88} (2013) 101301
  [arXiv:1310.1605 [astro-ph.CO]].
  
\bibitem{Carroll:2008br}
  S.~M.~Carroll, C.~Y.~Tseng and M.~B.~Wise,
  Phys.\ Rev.\ D {\bf 81} (2010) 083501
  [arXiv:0811.1086 [astro-ph]].

\bibitem{Blas:2014hya}
  D.~Blas, M.~Garny, T.~Konstandin and J.~Lesgourgues,
  ``Structure formation with massive neutrinos: going beyond linear theory,''
  arXiv:1408.2995 [astro-ph.CO].

\bibitem{Gubser:2004du}
  S.~S.~Gubser and P.~J.~E.~Peebles,
  Phys.\ Rev.\ D {\bf 70} (2004) 123511
  [hep-th/0407097].

\bibitem{Kesden:2006vz}
  M.~Kesden and M.~Kamionkowski,
  Phys.\ Rev.\ D {\bf 74} (2006) 083007
  [astro-ph/0608095].
  
\bibitem{Rocha:2012jg}
  M.~Rocha, A.~H.~G.~Peter, J.~S.~Bullock, M.~Kaplinghat, S.~Garrison-Kimmel, J.~Onorbe and L.~A.~Moustakas,
  Mon.\ Not.\ Roy.\ Astron.\ Soc.\  {\bf 430} (2013) 81
  [arXiv:1208.3025 [astro-ph.CO]].
  
\bibitem{Vogelsberger:2012ku}
  M.~Vogelsberger, J.~Zavala and A.~Loeb,
  Mon.\ Not.\ Roy.\ Astron.\ Soc.\  {\bf 423} (2012) 3740
  [arXiv:1201.5892 [astro-ph.CO]].

\bibitem{Kaplinghat:2013kqa}
  M.~Kaplinghat, S.~Tulin and H.~B.~Yu,
  arXiv:1308.0618 [hep-ph].

\end{thebibliography}
\end{document}